\documentclass[10pt, journal ]{IEEEtran}
\pdfoutput=1 

\bstctlcite{IEEEexample:BSTcontrol}
\usepackage{amsfonts}

\usepackage{epsfig,latexsym}
\usepackage{float}
\usepackage{indentfirst}
\usepackage{amsmath}
\usepackage{amssymb}
\usepackage{times}
\usepackage{subfigure}
\usepackage{stfloats,algorithm,algorithmic,bm}
\usepackage{psfrag}
\usepackage{cite}
\usepackage{subfigure}
\usepackage{stfloats}
\usepackage{bm}
\usepackage{amsthm}
\usepackage{color}

\usepackage{graphicx}

\usepackage{latexsym}

\usepackage{enumitem}
\newlist{Properties}{enumerate}{2}
\setlist[Properties]{label=Property \arabic*.,itemindent=*}

\usepackage[mathscr]{eucal}
\usepackage[center]{caption2}

\usepackage{array}

\usepackage{amsmath,amssymb}
\usepackage{bm} 

\usepackage{multirow}

\usepackage{longtable}
\usepackage{setspace}
\begin{document}
\title{Learning to Localize: A 3D CNN Approach to User Positioning in Massive MIMO-OFDM Systems}

\author{
	Chi~Wu,
	Xinping~Yi,
	Wenjin~Wang,
	Li~You,
	Qing~Huang,
	Xiqi~Gao,
	\thanks{
		C. Wu, W. Wang, L. You, Q. Huang and X. Q. Gao are with the National Mobile Communications Research Laboratory, Southeast University, Nanjing 210096, China (e-mail: chiwu@seu.edu.cn; wangwj@seu.edu.cn; liyou@seu.edu.cn; huangqing@seu.edu.cn; xqgao@seu.edu.cn).
	}
	\thanks{
		X. Yi is with the Department of Electrical Engineering and Electronics, University of Liverpool, L69 3BX, United Kingdom (email: xinping.yi@liverpool.ac.uk)
	}
	
}

\maketitle

\begin{abstract}
In this paper, we consider the user positioning problem in the massive multiple-input multiple-output (MIMO) orthogonal frequency-division multiplexing (OFDM) system with a uniform planner antenna (UPA) array.
Taking advantage of the UPA array geometry and wide bandwidth, we advocate the use of the angle-delay channel power matrix (ADCPM) as a new type of fingerprint to replace the traditional ones. 
The ADCPM embeds the stable and stationary multipath characteristics, e.g. delay, power, and angle in the vertical and horizontal directions, which are beneficial to positioning. 
Taking ADCPM fingerprints as the inputs, we propose a novel three-dimensional (3D) convolution neural network (CNN) enabled learning method to localize users' 3D positions. 
In particular, such a 3D CNN model consists of a convolution refinement module to refine the elementary feature maps from the ADCPM fingerprints, three extended Inception modules to extract the advanced feature maps, and a regression module to estimate the 3D positions. By intensive simulations, the proposed 3D CNN-enabled positioning method is demonstrated to achieve higher positioning accuracy than the traditional searching-based ones, with reduced computational complexity and storage overhead, and the ADCPM fingerprints  are more robust to noise contamination.
\end{abstract}

\section{Introduction}
As location-based applications (LBA) are extensively deployed in modern society, accurate positioning has received enormous attention in both industry and academia \cite{Emerging14Chin}. The global positioning system (GPS) has provided real-time outdoor positioning for the mobile terminal (MT), which can reach several meters of accuracy \cite{GPS2018}. However, in urban areas, the GPS positioning performance will be significantly degraded due to the blockage of buildings, cars, and pedestrians. 

Recently, user positioning by exploiting rich information of multipath wireless propagation has drawn a lot of attention. 
Various positioning methods have been proposed in the literature, including the geometry-based and the fingerprint-based positioning methods. The geometry-based positioning is a triangulating-to-localize method that relies on information of wireless signals from the users to the base stations (BSs), e.g., angle of arrival (AOA) \cite{Design17Bergen}, time of arrival (TOA) \cite{Novel09Xu}, time difference of arrival (TDOA) \cite{TDOA17Yuan}, and the received signal strength (RSS) \cite{Received18Cui}. However, the estimation errors of AOA/TOA/TDOA/RSS have an crucial influence on positioning accuracy in complex urban environments. In contrast, fingerprint-based positioning is a matching-to-localize method that consists of offline fingerprint database construction and online fingerprint matching and location prediction. In the offline phase, reference points (RPs) are selected, for which the pairs of fingerprints and corresponding positions are stored in the database. In the online phase, the position of the MT is estimated by searching the collected database and matching the input fingerprint with the stored ones. As an indicator of the surrounding environment, fingerprints have been widely adopted for user positioning in the complex multipath environment.

The most common feature used in fingerprint-based positioning is RSS \cite{Machine18Prasad,Cooperative19Wang,3DTarget17Tomic}, yet there are two shortcomings of the RSS-based fingerprint. On the one hand, RSS suffers from fast fading fluctuation and hardware heterogeneity and is therefore unstable for positioning. On the other hand, RSS only captures the coarsest channel information that cannot meet the demand in complex communication environment. Recently, some researchers proposed to use the channel state information (CSI) as the fingerprint \cite{ConFi17Chen,DeepFi2017Wang,Deep18Wu,CSIbased2018Decurninge}. Capturing more channel information than RSS, CSI has the potential to enhance the positioning accuracy. In certain scenarios (e.g., wireless sensor network \cite{DeepFi2017Wang}, and WiFi network \cite{ConFi17Chen}), however, due to limited bandwidth and number of antennas, the CSI fingerprint is insufficient to capture the multipath characteristics because of the low resolution in the spatial or frequency domain. 

Fortunately, such limitations can be overcome in massive multiple-input multiple-output orthogonal frequency-division multiplexing (MIMO-OFDM) system. Thanks to the large-scale antenna array and wide bandwidth, CSI fingerprints are able to capture rich multipath information including powers, angles, and delays \cite{Beam15Sun, Channel16You, Statistical16Li, Single18Sun, Spatial18Wang} for positioning. 
%
Various fingerprint-based positioning techniques have been proposed for massive MIMO/MIMO-OFDM systems. Of particular relevance is the approach proposed in \cite{Single18Sun}, where a weighted k-nearest neighbor (WKNN) algorithm is applied using the angle-delay domain channel information as fingerprints in the massive MIMO-OFDM system. Other approaches for positioning include the use of some sophisticated techniques, such as Gaussian processes regression in \cite{Machine18Prasad} and compressive sensing in \cite{Direct17Garcia}, to name a few. 


Most recently, deep learning has found application in user positioning, inspired by its great success in image recognition, speech signal processing, and self-driving. As a matter of fact, the fingerprint-based positioning can be cast into an image recognition problem, in which the fingerprints can be treated as images to recognize. For indoor positioning, a two-step training deep neural network (DNN) based positioning method  was proposed in \cite{OnDeep18Arnold} for the NLOS massive MIMO scenario, a single DNN classifier to determine the probabilities of the MT being on the collected RPs was used in \cite{Deep18Wu}, and a deep learning-based positioning method based on the classic deep belief nets (DBNs) with a stack of restricted Boltzmann machines (RBMs) was proposed in \cite{CSIBased17Wang}. For outdoor positioning, a deep convolution neural network (CNN) was utilized in \cite{Deep17Vieira} to map the CSI into the 2D position coordinates. 

However, the aforementioned methods were mainly dedicated to 2D positioning. 
When it comes to the UPA array, high angular resolution can be realized in both vertical and horizontal directions, which provides new opportunities for the fingerprint-based 3D positioning. 
To this end, we propose a novel deep learning based 3D positioning method for the MIMO-OFDM system with the UPA array, taking the angle-delay domain channel power matrix (ADCPM) as the fingerprint that contains multipath angles, delays, and powers information.
Instead of CSI fingerprints in the spatial-delay domain, we translate them into the angle-delay domain with a 3D discrete Fourier transform (DFT), by which the sparsity structures can be fully exploited.
%
%
To deal with the high dimensionality of the ADCPM, we propose a regression-oriented 3D CNN model that maps the ADCPM fingerprints into the 3D position coordinates directly. 
In particular, our 3D CNN model consists of four key components: (1) the convolution refinement module to refine the elementary feature maps from the ADCPM fingerprints; (2) 3D Inception blocks that are extended from the Inception blocks in AlexNet \cite{ImageNet12Alex} and GoogLeNet \cite{Going15Szegedy,Rethinking16Szegedy,Inceptionv416Szegedy}; (3) the average pooling to replace the full-connected layer at the bottom of the network, as suggested in \cite{NIN2019Lin}, to reduce the number of parameters of the network; (4) the batch normalization (BN) in each layer to improve the convergence speed and generalization ability of the network \cite{BN15Sergey}.
Such an end-to-end positioning method works in the following way.
In the offline training phase, the 3D CNN is trained by using the ADCPM fingerprints and the corresponding coordinates of RPs. In the online prediction phase, the trained 3D CNN is used to take new ADCPM fingerprints for position prediction. 

To summarize, our contributions are three-fold:

\begin{itemize}
	\item For the massive MIMO-OFDM systems with the UPA array, we propose a new type of fingerprints, ADCPM, which has rich and stable mutipath characteristics that are of close relevance to the position information.
	\item For the massive MIMO-OFDM systems with the UPA array, we propose a 3D CNN based positioning method, which achieves higher positioning accuracy with reduced storage and computational overhead than the searching-based methods (e.g., WKNN).
	\item We build a simulator and conduct extensive experiments to evaluate the proposed 3D CNN fingerprint-based positioning method with respect to positioning accuracy, storage and computation overhead.
\end{itemize}

The rest of the paper is organized as follows. In Section \ref{sec1}, we investigate the 3D MIMO-OFDM system and propose a new type of fingerprint extracted from the multipath channel characteristics. In Section \ref{sec2}, we introduce the design of our proposed 3D CNN-enabled positioning method,  followed by the detailed network architecture of our 3D CNN model in Section \ref{sec3}. Simulation results are presented in Section \ref{sec4}, and conclusion is given in Section \ref{sec5}.

{\bf Notations}: We use $\bar{\jmath} = \sqrt{-1}$ to denote the imaginary unit. Vectors and matrices are denoted in lower-case bold-faced characters and upper-case bold-faced characters respectively, the element indices of vector and matrix start with 0. We use $[\mathbf{a}]_i$, $[\mathbf{A}]_{i,j}$, $[\mathbf{A}]_i$ to denote the $i$th element of the vector $\mathbf{a}$, the $(i,j)$th element of matrix $\mathbf A$ and the $i$th column of matrix $\mathbf A$ respectively. The superscript $(\cdot)^T$, $(\cdot)^H$, $(\cdot)^*$ indicate the matrix transpose, conjugate-transpose and conjugate operation. The complex number field, real number field and, integer field are represented by $\mathbb{C}$, $\mathbb{R}$ and ${\mathbb{Z}}$. The symbol $ \odot $ denote the Kronecker product of two matrices. We use $E\{  \cdot \} $ to denote the expectation of random variable (RV) and random vector variable (RVC). $\left\lfloor x \right\rfloor $ denotes the largest integer not greater than $x$, ${\left\langle  \cdot  \right\rangle _N}$ denotes the modulo-$N$ operation. $\delta(\cdot)$ denotes the delta function.

\section{3D MIMO-OFDM System and Channel Characteristics}\label{sec1}
In this section, we start with the 3D massive MIMO-OFDM system modeling. The BS is equipped with a uniform planar array (UPA), comprising  $N$ antennas in each row and $M$ antennas in each column. Then we introduce the new type of fingerprint with the angle and delay information extracted from the channel characteristics.

\subsection{Channel model}
\begin{figure}[!htp]
	\centering
	\includegraphics[scale=0.6]{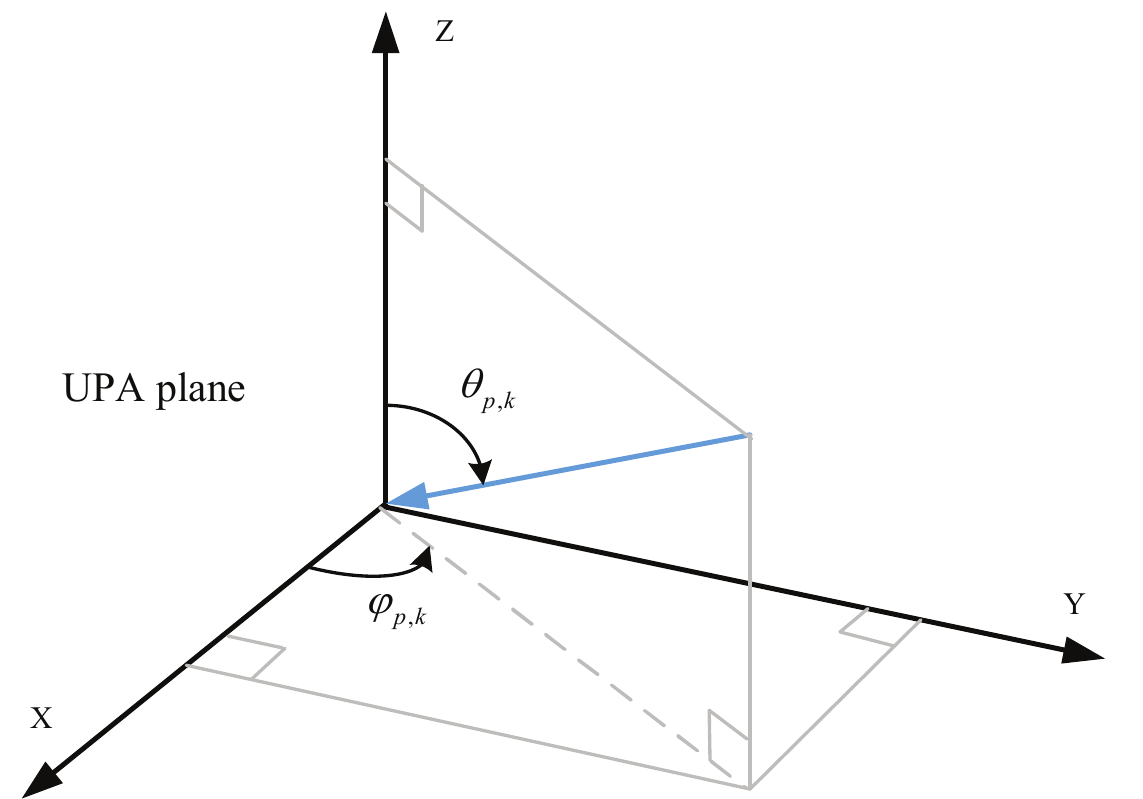}
	\caption{AOA of the received signal at BS antennas.}
	\label{3DAOA}
\end{figure}
We consider the uplink transmission in a wide-band massive MIMO wireless system. Owing to occlusions and reflections, wireless signals propagate through multipaths. Assume the number of multipaths is $N_p$ and the number of the MT is $K$, the AOA of the $p$th path for the $k$th MT can be decomposed into the elevation angle $0 \le {\theta _{p,k}} \le \pi $ in the vertical direction and the azimuth angle $0 \le {\varphi _{p,k}} \le \pi $ in the horizontal direction, as shown in Fig. \ref{3DAOA}. Thus, the array response vector ${\bf e}({\theta _{p,k}},{\varphi _{p,k}})$ can be written as
\begin{equation}\label{MIMO: eq2}
{\bf e}({\theta _{p,k}},{\varphi _{p,k}}) = {{\bf e}^{(v)}}({\theta _{p,k}}) \otimes {{\bf e}^{(h)}}({\theta _{p,k}},{\varphi _{p,k}}),
\end{equation}
with
\begin{equation}
{{\bf e}^{(v)}}({\theta _{p,k}}) = {[1,{e^{ - \bar{\jmath}2\pi \frac{{d_t^{(v)}}}{{{\lambda _c}}}\cos {\theta _{p,k}}}}, \cdots ,{e^{ - \bar{\jmath}2\pi (M - 1)\frac{{d_t^{(v)}}}{{{\lambda _c}}}\cos {\theta _{p,k}}}}]^T},
\end{equation}
and
\begin{equation}
\begin{aligned}
{{\bf e}^{(h)}}({\theta _{p,k}},{\varphi _{p,k}}) = &[1,{e^{ - \bar{\jmath}2\pi \frac{{d_t^{(h)}}}{{{\lambda _c}}}\sin {\theta _{p,k}}\cos {\varphi _{p,k}}}}, \cdots , \\ &{e^{ - \bar{\jmath}2\pi (N - 1)\frac{{d_t^{(h)}}}{{{\lambda _c}}}\sin {\theta _{p,k}}\cos {\varphi _{p,k}}}}]^T \hfill \\
\end{aligned},
\end{equation}
where $d_t^{(v)}$ and $d_t^{(h)}$ are the antenna spaces in the column and row respectively, and $\lambda_c$ is the carrier wavelength. Then, the channel impulse response (CIR) of the $p$th path for the $k$th MT is represented by
\begin{equation}\label{MIMO: eq1}
{{\bf{q}}_{p,k}} = {a_{p,k}}{\bf e}({\theta _{p,k}},{\varphi _{p,k}}).
\end{equation}

We consider OFDM modulation with $N_c$ sub-carriers, and the sample interval is $T_s$. We use $T_c=N_cT_s$ and $T_g=N_gT_s$ to denote the OFDM symbol duration and the cyclic prefix (a.k.a. guard interval) duration respectively. We assume the cyclic prefix duration $T_g$ is larger than the maximum channel delay of all the MTs. Let ${r_{p,k}} = \frac{{{\tau _{p,k}}}}{{{T_s}}}$, where ${\tau _{p,k}} = \frac{{{d_{p,k}}}}{c}$ is the time of arrival (TOA) of the $p$th path for the $k$th MT. The frequency of the $l$th sub-carrier is $f_l=\frac{l}{T_c}$. Thus, the channel frequency response (CFR) associated with the $k$th MT and the $l$th sub-carrier is written as
\begin{equation}
{{\mathbf{h}}_{k,l}} = \sum\limits_{p = 1}^{N_p} {{a_{p,k}}{\mathbf{e}}({\theta _{p,k}},{\varphi _{p,k}}){e^{ - \bar{\jmath}2\pi \frac{{l{r_{p,k}}}}{{{N_c}}}}}} .
\end{equation}
where ${a_{p,k}}\sim \mathcal{CN}(0,\sigma _{p,k}^2)$ is the complex path gain of the $p$th path.
The space-frequency channel response matrix (SFCRM) of the $k$th MT known to the BS can be denoted by the concatenation of ${{\mathbf{h}}_{k,l}} $, i.e.
\begin{equation}
{{\mathbf{H}}_k} = [{{\mathbf{h}}_{k,0}},{{\mathbf{h}}_{k,1}}, \cdots ,{{\mathbf{h}}_{k,{N_c} - 1}}].
\end{equation}

\subsection{Fingerprint from Angle-Delay Domain} \label{fingerprint extracted}

Define ${{\bf{V}}_M}\in {\mathbb{C}^{{M} \times {M}}}$ and ${{\bf{V}}_N}\in {\mathbb{C}^{{N} \times {N}}}$ as phase-shifted discrete Fourier transform (DFT) matrices with the $(m,n)$th element ${\left[ {{{\bf{V}}_M}} \right]_{m,n}} = \frac{1}{{\sqrt M }}{e^{ - \bar{\jmath}2\pi \frac{{m(n - M/2)}}{M}}}$ and ${\left[ {{{\bf{V}}_N}} \right]_{m,n}} = \frac{1}{{\sqrt N }}{e^{ - \bar{\jmath}2\pi \frac{{m(n - N/2)}}{N}}}$ respectively.
Define ${{\mathbf{F}}_{{N_c} \times {N_g}}} \in {\mathbb{C}^{{N_c} \times {N_g}}}$ as the matrix composed of the first $N_g$ columns of DFT matrix $\mathbf{F}_{N_c}$ with the $(i,j)$th element ${[{{\mathbf{F}}_{{N_c} \times {N_g}}}]_{i,j}} \triangleq \frac{1}{{\sqrt {{N_c}} }}{e^{ - \bar{\jmath}2\pi \frac{{ij}}{{{N_c}}}}}$.

The fingerprints used in positioning are required to be closely linked to the MT's positions. The CFR describes the space-frequency domain characteristics, but it is hard to build an intuitive relationship between position and CFR. Due to complex and changeable multipath propagation in the wireless channel, the AOAs and TOAs of received signals are unique for different positions. Therefore, it is sufficient to extract a fingerprint from the angle-delay domain. We reconstruct CFR matrix into angle-delay domain matrix, and ${{\mathbf{G}}_k}$ is referred to as the angle-delay domain channel response matrix (ADCRM) of the $k$th MT, given by
\begin{equation}\label{MIMO: eq26}
{{\mathbf{G}}_k} = \frac{1}{\sqrt {MN{N_c}}}\left( {{\mathbf{V}}_M^H \otimes {\mathbf{V}}_N^H} \right){{\mathbf{H}}_k}{\mathbf{F}}_{{N_c} \times {N_g}}^*.
\end{equation}
As such, the angle-delay domain channel power matrix (ADCPM) of the $k$th MT is introduced and used as a fingerprint hereafter, i.e.,
\begin{equation}\label{MIMO: eq29}
{{\mathbf{\Omega }}_k} \triangleq E\left\{ {{{\mathbf{G}}_k} \odot {\mathbf{G}}_k^*} \right\},
\end{equation}
with
\begin{equation}\label{MIMO: eq31}
{[{{\mathbf{\Omega }}_k}]_{i,j}} \triangleq E\left\{ {{{\left| {{{[{{\mathbf{G}}_k}]}_{i,j}}} \right|}^2}} \right\}.
\end{equation}

Define the following function:
\begin{equation}\label{MIMO: fM}
f_{M}(x) = \frac{{\sin (Mx)}}{{M\sin (x)}}.
\end{equation}

\textit{Theorem 1:}
 For 3D MIMO-OFDM systems with UPA at the BS, when $M\to \infty $ and $N_c\to \infty$, the ADCPM is concentrated on specific position in the vertical angle-delay domain, given by
\begin{equation}\label{MIMO: eq34}
\begin{aligned}
\mathop {\lim }\limits_{M \to \infty,{N_c} \to \infty } [{{\mathbf{\Omega }}_k}]_{i,j} = &\sum\limits_{p = 1}^{N_p} \sigma _{p,k}^2
	 f_N^2\left( {\frac{{{{\bar n}_{p,k}} - {\left\langle i \right\rangle }_N  }}{{2N}}} \right) \\&\cdot \delta \left( {{\left\lfloor {i/N} \right\rfloor} - {{\bar m}_{p,k}}} \right) \delta \left( {j - {r_{p,k}}} \right)
\end{aligned}.
\end{equation}
where 
\begin{equation}
{{\bar m}_{p,k}} = \frac{M}{2} + \frac{{Md_t^{(v)}}}{{{\lambda _c}}}\cos {\theta _{p,k}},
\end{equation}
When $N\to \infty$ and $N_c\to \infty$, the ADCPM has the similar conclusion in the horizontal angle-delay domain, given by
\begin{equation}\label{MIMO: eq35}
\begin{aligned}
\mathop {\lim }\limits_{N \to \infty ,{N_c} \to \infty } [{{\mathbf{\Omega }}_k}]_{i,j} = &\sum\limits_{p = 1}^{N_p} \sigma _{p,k}^2f_M^2\left( {\frac{{{{\bar m}_{p,k}} - { \left\lfloor {i/N} \right\rfloor }}}{{2M}}} \right)\\
&\cdot \delta \left( {  {\left\langle i \right\rangle }_N - {{\bar n}_{p,k}}} \right) \delta \left( {j - {r_{p,k}}} \right)
\end{aligned}.
\end{equation}
where
\begin{equation}
{{\bar n}_{p,k}} = \frac{N}{2} + \frac{{Nd_t^{(h)}}}{{{\lambda _c}}}\sin {\theta _{p,k}}\cos {\varphi _{p,k}}.
\end{equation}
When $M\to \infty $, $N\to \infty$ and $N_c\to \infty$, the ADCPM is concentrated on the $({{\bar m}_{p,k}}N + {{\bar n}_{p,k}})$th angle direction and the $r_{p,k}$th delay direction, given by
\begin{equation}\label{MIMO: eq36}
\begin{aligned}
\mathop {\lim }\limits_{M \to \infty ,N \to \infty ,{N_c} \to \infty } [{{\mathbf{\Omega }}_k}]_{i,j} = &\sum\limits_{p = 1}^{N_p} \sigma _{p,k}^2  \delta \left( {i - {{\bar m}_{p,k}} N - {{\bar n}_{p,k}}} \right)\\
&\cdot\delta \left( {j - {r_{p,k}}} \right)
\end{aligned}.
\end{equation}

\textit{Proof:} See appendix \ref{proof of thro2}. \qed

\textit{Remark 1:} For 3D massive MIMO-OFDM systems, by (\ref{MIMO: eq26}) and (\ref{MIMO: eq29}), the CFR in the space-frequency domain is translated into the ADCPM in the angle-delay domain. Theorem 1 reveals that the $(i,j)$th element in the ADCPM corresponds to the average channel power of the $i$th AOA and the $j$th TOA. 

\textit{Remark 2:} In Theorem 1, $\delta(\cdot)$ determines the sparsity and specifies the sparsity pattern.
When $N_c\to \infty$, $M\to \infty $ and/or $N\to \infty$, the ADCPM is asymptotically a sparse matrix in the sense that most elements are equal to zero. For the finite $N_c$, $M$ and $N$, it will be shown later by numerical results that the sparsity maintains.

\textit{Remark 3:} From Theorem 1, the sparsity pattern of the ADCPM depends on both the AoAs and the TOAs of multiple paths.
For two MTs located at different positions, it is unlikely all the multipath components of the signals are identical, so are the corresponding ADCPMs. As such, the ADCPM can be a unique indicator to discriminate MTs from different geographical positions. 

According to Theorem 1, the ADCPM is suitable to serve as the fingerprint for positioning, as the ADCPMs meet the following requirements for fingerprints.
\begin{enumerate}
\item The ADCPMs are closely related to geographic locations. As stated above, the ADCPM embeds information of the AOAs, the TOAs, and the channel power corresponding to a specific geographic location.
\item The ADCPMs have a sufficient degree of discrimination between different geographical locations, which increases as two locations are farther apart. 
\item The ADCPM is stationary in the sense that it keeps unchanged over a relatively long period for a given location. In the multipath environment, as long as the distribution of the scatterers does not change, the angle and delay keep unchanged, so does the ADCPM.
\end{enumerate}

In addition, the ADCPM is convenient to be extracted from the channel state information (CSI) at the BS through wideband signal processing. To conclude, the ADCPM provides a highly differentiable, stable, and easily accessible indicator for different geographic location, and therefore is an ideal fingerprint for positioning.

\section{Convolution Neural Network for Positioning}\label{sec2}
Provided the ADCPM as the fingerprint in Section \ref{fingerprint extracted}, the problem arises as to how to realize positioning by exploiting the structural properties of ADCPM. 

As the ADCPM fingerprint can be seen as an image, the widely used end-to-end image recognition method - convolution neural network (CNN) empowered deep learning - can be applied here for positioning. Thanks to its characteristics of space invariant, parameter sharing, and hierarchical representations, CNN is more efficient than the traditional fully-connected networks when dealing with large dimensional inputs \cite{Deep16Goodfellow}. For 3D massive MIMO-OFDM systems, the high dimensional ADCPM has sparsity patterns, which suggests that CNN could be an ideal positioning method to extract the positioning-related features from the ADCPM and convert these features into position information with relatively low computational complexity.

\subsection{The Sparse ADCPM as Input}
\begin{figure}[!htp]
	\centering
	\includegraphics[scale=0.6]{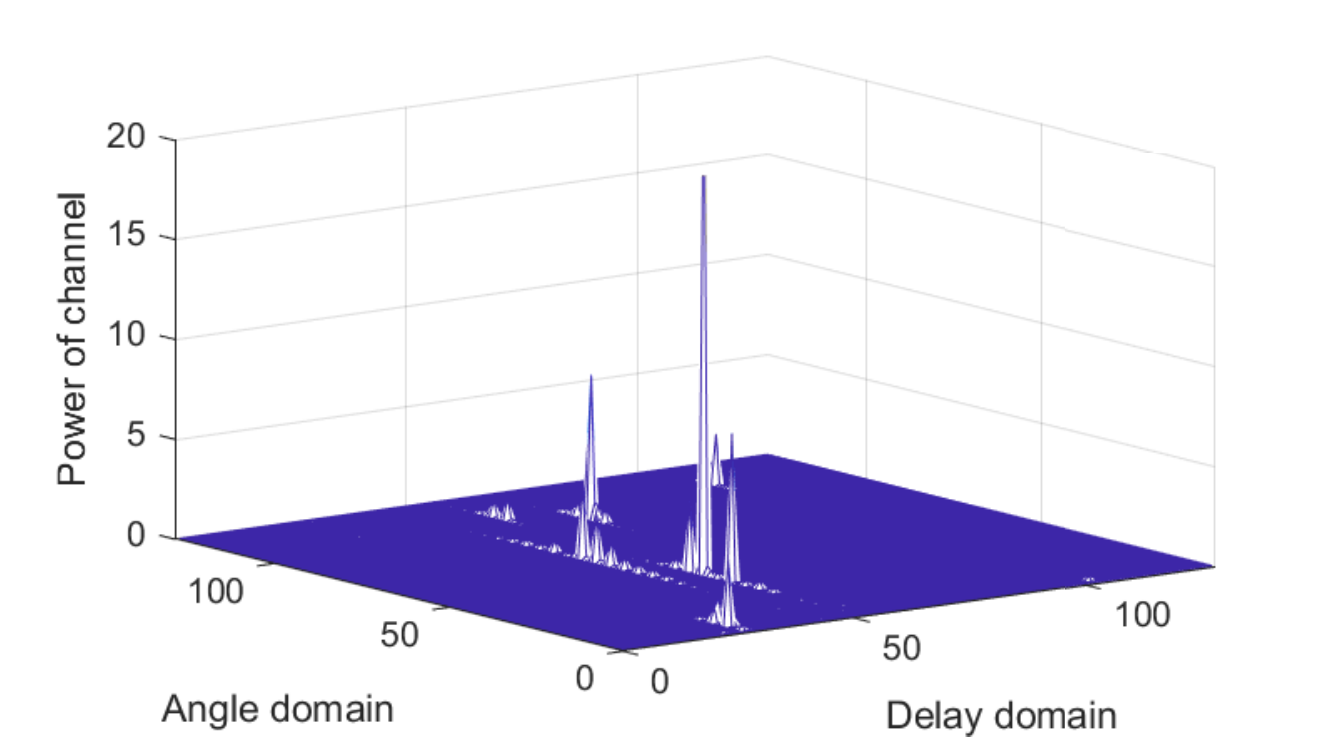}
	\caption{The extracted ADCPM fingerprint.}
	\label{ADCPM_2005}
\end{figure}

As shown in Fig. \ref{ADCPM_2005}, the asymptotic property  in Theorem 1 of the ADCPM sparsity pattern in the angle-delay domain still maintains in the practical setting with a finite number of antennas and limited bandwidth.
Due to the sparsity pattern, the ADCPM fingerprint makes the difference of the channel at different positions more distinguishable. As such the features of the channel are easier to be extracted by a neural network.

In fact, the feature maps in the higher layers of CNN are also sparse because they solely focus on the discriminant structure within the input picture \cite{Visualizing14Zeiler}. 
By using the sparse ADCPM as the input, it is easier for the neural network to capture the characteristic information of the channel, thereby simplifying the neural network structure and speeding up the convergence of the neural network.

In Theorem 1, the rows and the columns of the ADCPM corresponds respectively to the angles and the delay domain. Noticing that the angle in the 3D space can be described by a pair of vertical and horizontal angles, we reshape the ADCPM into a three-dimensional tensor in the following way
\begin{equation}
\left[\mathbf{X}_{k}\right]_{m, n, j}=\left[\boldsymbol{\Omega}_{k}\right]_{mN+n, j},
\end{equation}
where ${{\mathbf{{ X}}}_k} \in {\mathbb{R}^{M \times N \times {N_g}}}$ is the 3D ADCPM of the $k$th MT with three dimensions indicating vertical angle, horizontal angle, and delay respectively. We use hereafter the reshaped ADCPM as the input of neural networks.

\subsection{Regression-oriented Positioning}
The CNN-based regression-oriented positioning is actually a complex non-linear function.
Denote by $f(\cdot)$ such a function that predicts the true position ${\bf a}_k=(x,y,z)$ according to the user's 3D ADCPM ${\bf X}_k$, that is,
\begin{equation}
{{{\mathbf{\hat a}}}_k} = f({{\mathbf{X }}_k}),
\end{equation}
where ${{{\mathbf{\hat a}}}_k}=(\hat{x}, \hat{y}, \hat{z})$ is the prediction by the CNN.
We use regression analysis to find the mapping function through minimizing the localization error between the true coordinate and the prediction
\begin{equation}
{e_k} = ||{{{\mathbf{\hat a}}}_k} - {{{\mathbf{ a}}}_k}||.
\end{equation}

Note that the traditional CNN models are often used for image classification with the last layer activated by a softmax function. As a matter of fact, the CNN model itself can be regarded as a regression function if the softmax function is replaced with a fully connected layer without activation function. If we use ${{\bf{T}}_0}$ to denote the output before the last layer, then the estimated position after the last layer can be written by
\begin{equation}\label{eq19}
{\bf{\hat a}} = {{\bf{W }}}{\mathop{\rm vec}\nolimits} \left\{ {{{\bf{T}}_0}} \right\} + {{\bf{ b}}},
\end{equation} 
where ${{\bf{W }}}$ and ${{\bf{ b}}}$ are the parametric weight matrix and bias vector respectively that can be learned together with the training of previous CNN layers.

Let ${\boldsymbol \theta} $ be the set of trainable parameters of the neural network. The cost function with respect to the mean square error (MSE) of the training data set can be given by
\begin{equation}\label{eq20}
J(\boldsymbol{\theta})=\frac{1}{N_{\text {train}}} \sum_{i=1}^{N_{\text {train}}}\left\|\mathbf{a}_{i}-\hat{\mathbf{a}}_{i}\right\|^{2}+\frac{\lambda}{2} \boldsymbol{\theta}^{T} \boldsymbol{\theta},
\end{equation}
where $N_{\text{train}}$ is the number of training samples, the second term employs the $L_2$ regularization to avoid over-fitting, and $\lambda$ is the weight factor of the $L_2$ regularization.


\section{Proposed 3D CNN Structure}\label{sec3}
\begin{figure*}[htbp]
	\centering
	\includegraphics[width=0.8\textwidth]{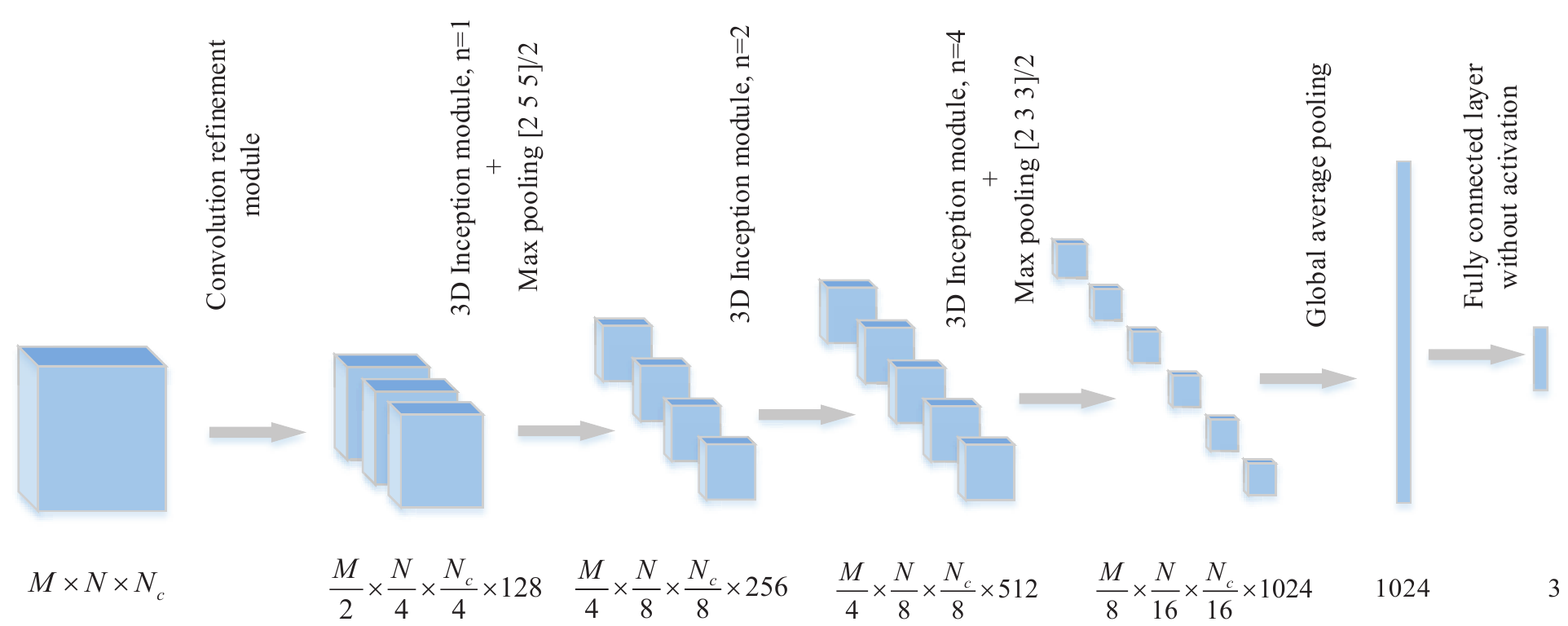}
	\caption{The network architecture of the proposed 3D CNN for fingerprint-based positioning.}
	\label{Proposed3DCNN}
\end{figure*}

Inspired by the 3D structure of ${{\mathbf{X }}_k}$, we propose to use the 3D CNN to realize the mapping function. Fig. \ref{Proposed3DCNN} shows the network architecture of the proposed 3D CNN for fingerprint-based positioning, which is composed of a convolution refinement module, three extended 3D Inception modules, and a regression module. In addition, the max pooling is used for downsampling, and its description parameters are presented in the form of (size/stride). The convolution refinement module first refines the elementary feature maps from the 3D ADCPM. Then we modify and extend the Inception module into a 3D form to extract the advanced feature maps. Further, the regression module estimates the 3D position by employing a global average pooling layer and a fully connected layer without activation function.

Before proceeding further, we introduce the 3D convolution-normalization-activation (CNA) layer, an elementary building block used in our 3D CNN. The 3D CNA layer consists of three parts: a 3D convolution, a BN transform, and an activation function. Consider the input feature maps ${\mathbf{I}} \in {\mathbb{R}^{H \times W \times L \times P}}$, where $H$, $W$, $L$, $P$ denote the height, width, length, and the number of channels of ${\mathbf{I}}$ respectively. Given the 3D convolution kernel $\mathbf{K} \in \mathbb{R}^{K_{1} \times K_{2} \times K_{3} \times P \times Q}$, where $K_1$, $K_2$, $K_3$ are the kernel size of the height, width, and length respectively, and $Q$ is the number of output channels. The convolution is performed with zero-padding \cite{Deep16Goodfellow} and the strides are set to be 1. By convolving the 3D kernel with the input feature maps, the 3D convolution yields output $\mathbf{O} \in \mathbb{R}^{H \times W \times L \times Q}$ with
\begin{equation}\label{MIMO: eq37}
[\mathbf{O}]_{m, n, j, q}=\sum_{k_{1}, k_{2}, k_{3}} \sum_{p}[\mathbf{K}]_{k_{1}, k_{2}, k_{3}, p, q} \cdot[\mathbf{I}]_{m+k_{1}, n+k_{2}, j+k_{3}, p}.
\end{equation}
Right after the 3D convolution, we adopt the BN to improve the convergence rate and generalization performance \cite{BN15Sergey}, followed by the rectified linear unit (ReLU) as the activation function. That is, the output feature maps of the 3D CNA layer $\mathbf{T} \in \mathbb{R}^{H \times W \times L \times Q}$ are given by

\begin{equation}
{[{\mathbf{T}}]_{m,n,j,q}} = \max \left( {0,{\text{BN}}\left( {{{[{\mathbf{O}}]}_{m,n,j,q}}} \right)} \right),
\end{equation}
where $\text{BN}(\cdot)$ represents the BN transform. Note that we discard the bias term in (\ref{MIMO: eq37}) as suggested in \cite{BN15Sergey}.

In what follows, we describe the individual modules and the overall network structure in detail.

\subsection{Convolution Refinement Module} 
The convolution refinement module is designed to extract features from the input 3D ADCPM. When designing the module, we take into account the corresponding physical meaning of each dimension of the 3D ADCPM. While the existing designs use the symmetric kernel, for which the size of each dimension is equal, we propose to use the asymmetric convolution kernel, based on the intuition that the size of kernel should reflect the correlation statistics of the corresponding dimension.
As a consequence, the size of the subsequent max pooling should also change accordingly. From Fig. \ref{ADCPM_2005}, we observe that the input units are more concentrated in the angular dimension and relatively dispersed in the time delay dimension. It suggests that the size of the corresponding convolution kernel size of the delay dimension can be larger than that of the angle dimension.
At the same time, it is worth noting that the vertical and horizontal angles are somehow correlated, while the delay dimension is independent of the other two. Therefore, the delay-vertical angle and the delay-horizontal angle domains demand special treatments, compared with the vertical-horizontal angle domain. This motivates our design of the convolution refinement module.

\begin{figure}[!htp]
	\centering
	\includegraphics[width=0.45\textwidth]{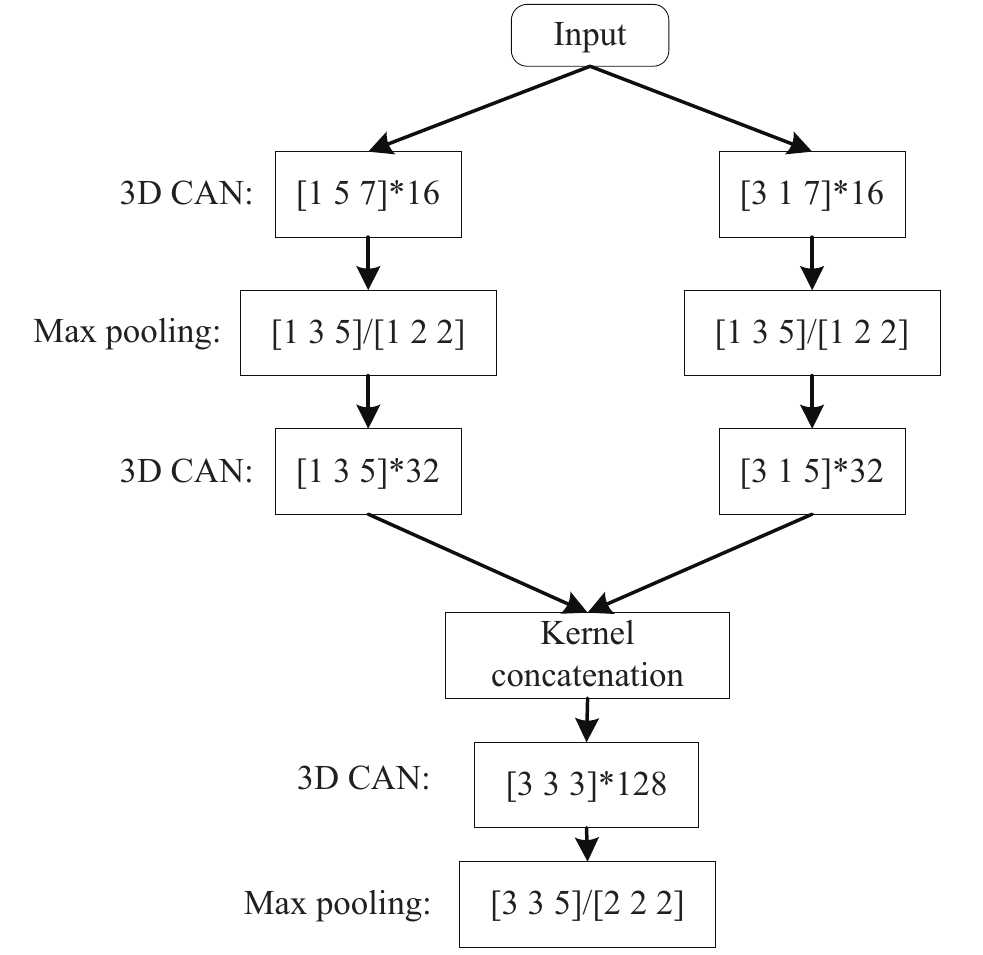}
	\caption{The proposed convolution refinement module with two branches of CAN layers taking special case of the delay-vertical angle and delay-horizontal angle domains.}
	\label{ConvolutionRefinementModule}
\end{figure}

We illustrate the case of $M=8$, $N=16$, and $N_g=128$ as an example to describe the convolution refinement module. The structure of the convolution refinement module is shown in Fig. \ref{ConvolutionRefinementModule}. The expression in the box is in the form of (size*number/stride). As one of the key designs, we build two branches in this module, each of which consists of two 3D CNA layers and one max pooling layer. The left branch aims to extract the information from the delay-horizontal angle domain, while the right branch is dedicated to the information from the delay-vertical angle domain. We emphasize here that both branches use asymmetrical convolution kernels and pooling size. The outputs of two branches are combined using kernel-wise concatenation. After that, we employ a 3D CNA layer with the symmetrical kernel to extract the information from all three dimensions, followed by a max pooling to downsample the feature maps.

\subsection{3D Inception Module} 
\begin{figure}[t]
	\centering
	\includegraphics[width=0.45\textwidth]{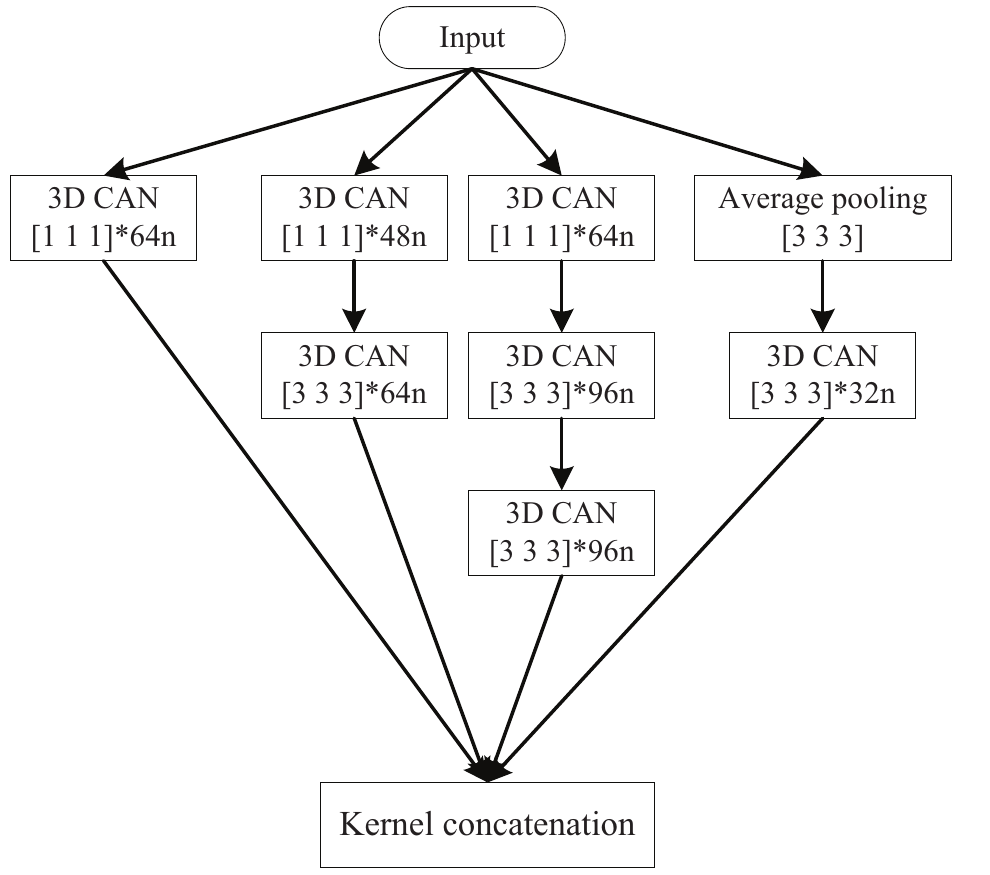}
	\caption{The modified 3D Inception module.}
	\label{3DInceptionModule}
\end{figure}

In the deeper layers, we use the 3D Inception module to extract more precise features. The Inception module is a combination of 2D convolutions with different kernel sizes, e.g. $1 \times 1$, $3 \times 3$, $5 \times 5$, with a parallel average-pooling operation \cite{Going15Szegedy}. We extend the Inception module into a 3D form by replacing the 2D convolution with the 3D convolution. The structure has four parallel branches, whose outputs are concatenated into a single output vector that servers as the input of the next stage, as shown in Fig. \ref{3DInceptionModule}, where $n$ is the factor of kernel number. The $5 \times 5 \times 5$ kernel is replaced by two cascaded $3\times 3 \times 3$ kernels to reduce the computational overhead, as in \cite{Going15Szegedy}.


Since the correlation statistics in the deeper layers are unknown, the usage of 3D convolution kernels with different sizes can avoid the loss of important features.

\subsection{Regression Module}
The regression module includes a global average pooling and a fully connected layer without activation function. While existing designs tend to add several layers of fully connected layers before the output layer of the network, we use a global average pooling to replace them. The global average pooling takes the average of each feature map \cite{NIN2019Lin}, and is used to reshape the output of the convolution layers for the final fully connected layer. By replacing the middle fully connected layers with the global average pooling, we can avoid the huge parameters brought by the middle fully connected layers. Thus, the global pooling improves convergence rate while reducing overfitting.
The output of global average pooling in vector form is fed into the fully connected layer indicated by (\ref{eq19}) directly. The output of the regression module is the estimated coordinates of the position.

\section{Simulation Results}\label{sec4}
In this section, we first introduce the simulation setup. Then we evaluate the positioning accuracy, time and storage overhead of the proposed 3D CNN-enabled positioning method. Next, we demonstrate the influence of antenna and bandwidth to our proposed 3D CNN-enabled positioning method. At last, we evaluate the robustness of our positioning method with the noise contaminated inputs.

\subsection{Simulation Setup}
To simulate the positioning process, we implement a spatial consistency channel model based on the QuaDriGa model \cite{Jaeckel14QuaDRiGa}. Spatial consistency means the similarity of channels at two adjacent positions caused by similar scatterer environment. The fingerprint-based positioning relies on the similarity between fingerprints obtained from the channel. Thus, it is critical for our simulator to contain spatial consistency. The QuaDriGa model incorporates time evolution to realize spatial consistency in a preset track. Based on the transition idea from the WINNER model \cite{Jaeckel14QuaDRiGa}, we extend the spatial consistency into the whole 3D space by using reference points (RPs) transition. The RPs are uniformly distributed in the 3D space, and the interval between them is equal to the correlation distance used to describe the correlation of large-scale parameters (LSPs) \cite{WINNERReport}. A stochastic part generates the channel coefficients of the RPs, following \cite{3GPP36873, 3GPP38901}. Then based on the geographic position, the channel of arbitrary MT in the 3D space can be generated by the transitions between the RPs. Thus, we build a simulation environment including the geographical correlation between the fingerprints of two neighboring positions.

In our simulation, we consider the 3GPP urban macro (UMa) NLOS scenario, where the carrier frequency is set to 2GHz. We choose the NLOS scenario instead of the LOS one because the positioning in the NLOS scenario is more challenging due to a lot of occlusions and reflections. The size of the positioning area is 30m $\times$ 30m $\times$ 9m, with the center point of its bottom surface coincides with the origin. The BS is equipped with a UPA at $(-100,0,25)$m, the antenna plane is perpendicular to the ground, facing the positioning area. The MT is equipped with an Omni-directional antenna. 
Unless otherwise specified, the transmission bandwidth is 20MHz, and the configuration of UPA is $M=8$, $N=16$. For simplicity but without loss of generality, we divide the cube positioning area into three planes with heights 1.5m, 4.5m, and 7.5m respectively. In the offline phase, the training points are uniformly selected on these three planes with an interval of 1m, then the fingerprints and the corresponding 3D positions are collected. In the online phase, the test points are randomly distributed on these three planes with a total number of 1000, for which their positions are inferred by putting the fingerprint into the trained 3D CNN.

The fingerprints and the corresponding 3D positions of the training points are generated and saved using MATLAB 2018b, and the 3D CNN training and testing are processed using TensorFlow 1.9. Our simulation is carried out on a workstation equipped with two E5 2643v3 CPUs and one Titan X Pascal 12GB GPU. The time overhead mentioned below only refers to the run time on TensorFlow 1.9.

\subsection{Comparison with Other Positioning Algorithms}

To evaluate the performance of the proposed 3D CNN-enabled positioning method, we use the WKNN positioning method proposed in \cite{Fingerprint17Sun} as the benchmark. For fair comparison, we modify the fingerprint similarity criterion by adding the normalization, given by
\begin{equation}\label{MIMO: eqCMD}
J\left(\boldsymbol{\Omega}_{k}, \boldsymbol{\Omega}_{l}\right)=\frac{\operatorname{Tr}\left(\boldsymbol{\Omega}_{k}^{T} \boldsymbol{\Omega}_{l}\right)}{\left\|\boldsymbol{\Omega}_{k}\right\|_{F}\left\|\boldsymbol{\Omega}_{l}\right\|_{F}},
\end{equation}
where ${\left\|  \cdot  \right\|_F}$ denotes the Frobenius norm, and ${\text{Tr}}( \cdot )$ denotes the ``trace" operator. The modification is inspired by the correlation matrix distance (CMD) proposed in \cite{Correlation05Herdin} to measure the spatial structure of the non-stationary MIMO channel. The normalization term confines the similarity of fingerprints between 0 and 1 to eliminate the influence introduced by different channel powers. Thus the normalized fingerprint similarity criterion (\ref{MIMO: eqCMD}) makes the WKNN positioning algorithm more accurate, which ensures the fairness of comparison. Then, the number of neighbors $K=4$ is adopted in our simulation.

On the other hand, to demonstrate the benefit of the 3D CNN model, we also make the comparison with a downgraded 2D CNN model. We propose a 2D CNN-enabled positioning method with the ADCPM presented by (\ref{MIMO: eq29}) as the input. In that case, the ADCPM is regarded as an image with sparse highlights (supports), where the vertical and horizontal angles are collocated in the same dimension. Table \ref{2DCNN} specifies the structure of the proposed 2D CNN where the 2D Inception module is modified from the 3D Inception module in Fig. \ref{3DInceptionModule} by replacing the 3D convolution with the 2D convolution.
\begin{table}[t]
	\centering 
	\caption{2D CNN parameters} \label{2DCNN}
	\vspace{0.5em}
	\begin{tabular}{c|c}
		\hline \hline
		Module & size*number/stride  \\
		\hline
		CNA & [15 15]*32  \\
		\hline
	    Max pooling & [5 5]/2  \\
		\hline
		CNA & [7 7]*64  \\
		\hline
		CNA & [5 5]*128 \\
		\hline
		Max pooling & [5 5]/2\\
		\hline
		2D Inception module & $n=1$   \\
		\hline
		Max pooling & [5 5]/2 \\
		\hline
		2D Inception module & $n=2$ \\
		\hline
		2D Inception module & $n=4$ \\
		\hline
		Max pooling & [3 3]/2  \\
		\hline
		Global average pooling & [8 8] \\
		\hline
		Fully connected layer & 1024*3 \\
\hline
		\hline
	\end{tabular}
\end{table}

\begin{figure}[t]
	\centering
	\includegraphics[scale=0.6]{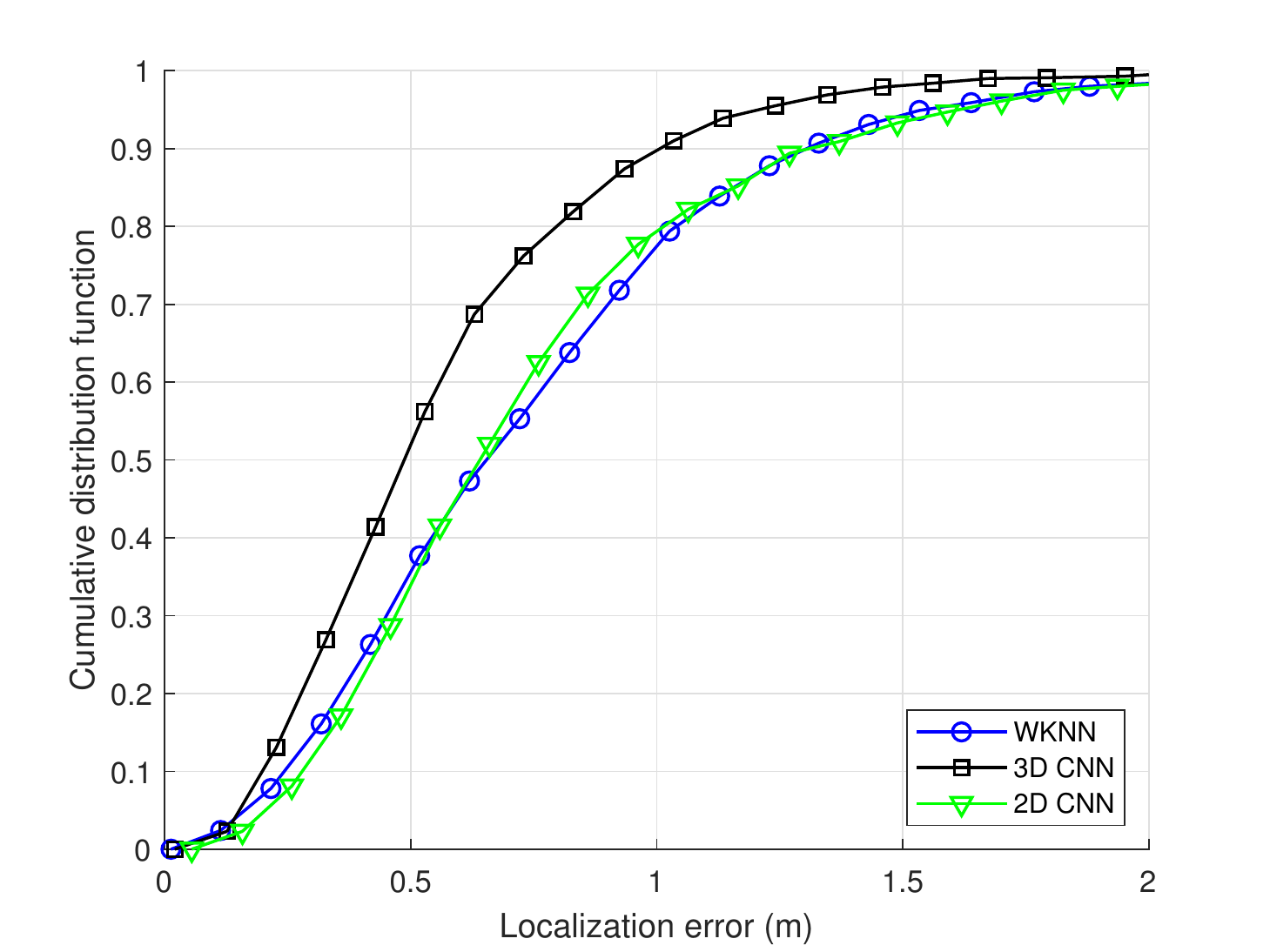}
	\caption{The CDF of localization error with different positioning methods.}
	\label{errorCDFViaMethod}
\end{figure}

\begin{figure}[t]
	\centering
	\includegraphics[scale=0.6]{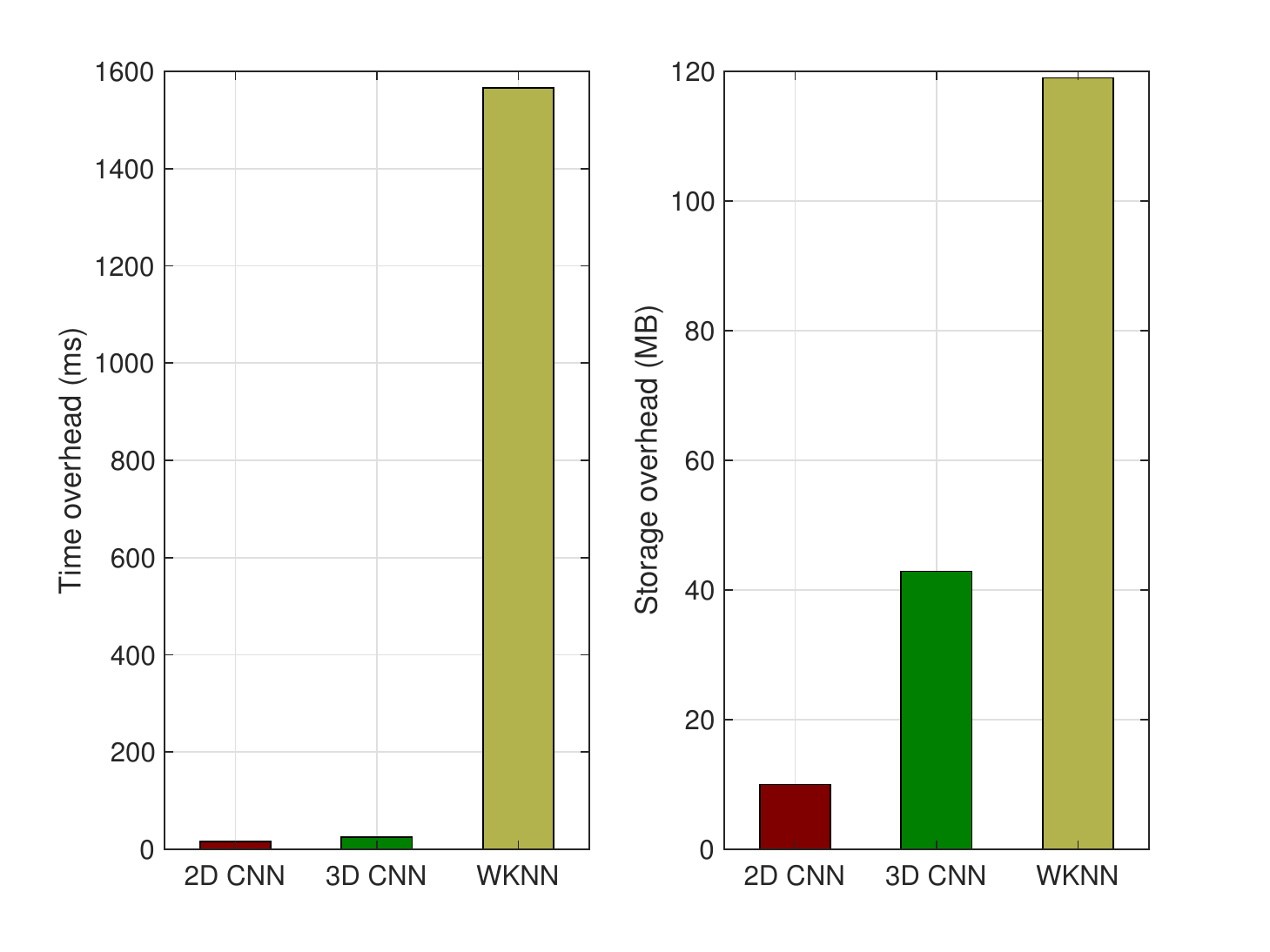}
	\caption{The time overhead (left) and the storage overhead (right) with different positioning
		method.}
	\label{Time_Storage}
\end{figure}

We first compare the positioning accuracy of the three methods. The cumulative distribution function (CDF) of the localization error using different methods is illustrated in Fig. \ref{errorCDFViaMethod}. The 3D CNN-enabled positioning method reaches the highest positioning accuracy with 90\% of localization errors within 1m. Compared with the 3D counterpart, the 2D CNN method realizes the inferior positioning accuracy with only 80\%, but is still superior to the WKNN positioning method with 78\%.

The run time overhead is defined as the time spent per user positioning in the online phase, and we employ it to measure the computational complexity. The storage overhead refers to the required storage resource for positioning. We compare the time  and storage overhead of the three positioning methods in Fig. \ref{Time_Storage}. The 2D CNN-enabled positioning method requires the least time overhead, e.g. 15.78ms, and the least storage overhead, e.g. 10MB. Compared with the 2D counterpart, 3D CNN gains higher positioning accuracy at the cost of higher time overhead (24.42ms running time) and storage overhead (42.9MB). The WKNN positioning method has the highest computational complexity (1566ms) and storage requirement (119MB), as it needs to store the database collected in the positioning area and then search through the positioning area to find the nearest positions. Thus the time and storage overhead of WKNN positioning method will increase with the expansion of the positioning area. In contrast, the 2D/3D CNN-enabled positioning methods only need to store the trained parameters and infer the position directly through the trained regression networks, by which the time and storage overhead is independent of the size of the positioning area.

In conclusion, the proposed 3D CNN-enabled positioning method outperforms the WKNN positioning method in terms of positioning accuracy, time overhead and storage overhead. The proposed 2D CNN-enabled positioning method can be regarded as a simplification of the 3D CNN-enabled positioning method, which reduces the time overhead and storage overhead at the sacrifice of the positioning accuracy.
\subsection{Different Configurations of Antenna Array and Bandwidth}
\begin{figure}[t]
	\centering
	\includegraphics[scale=0.6]{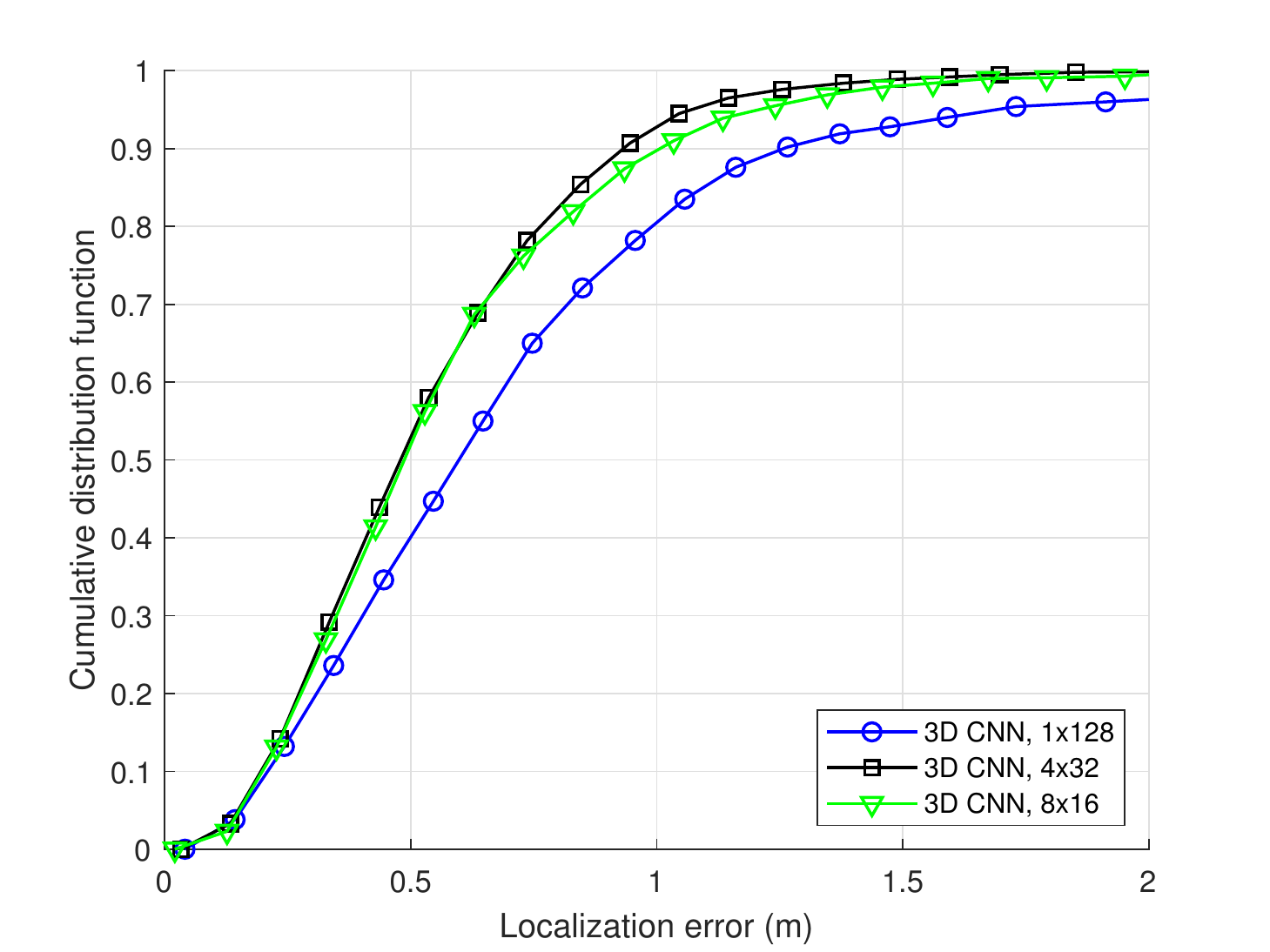}
	\caption{The CDF of localization error with different number of BS antennas in the column and row.}
	\label{3DCNNerrorViaAnt}
\end{figure}
\begin{figure}[!htp]
	\centering
	\includegraphics[scale=0.6]{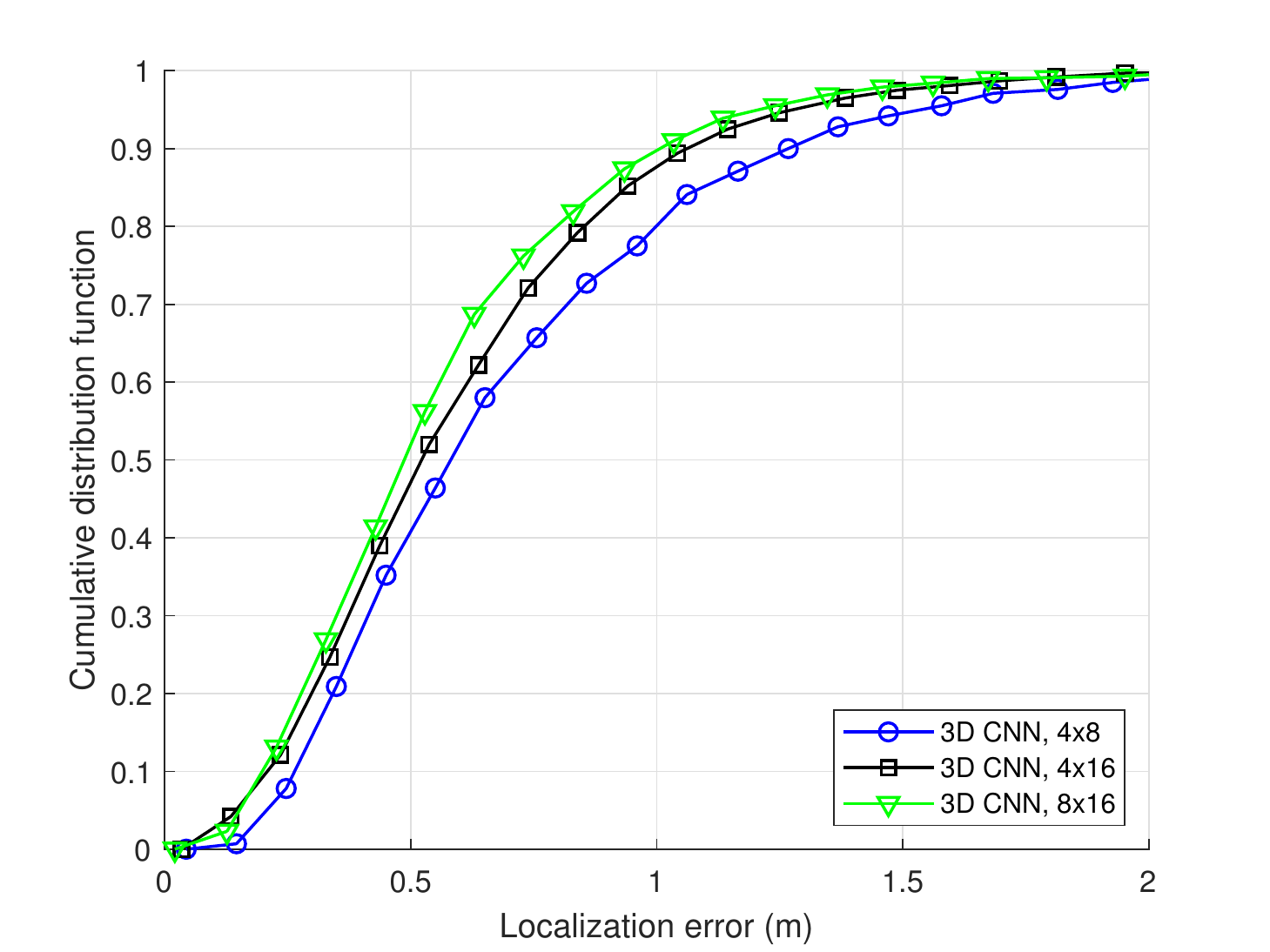}
	\caption{The CDF of localization error with different number of BS antennas.}
	\label{3DCNNerrorViaAntNum}
\end{figure}

\begin{figure}[t]
	\centering
	\includegraphics[scale=0.6]{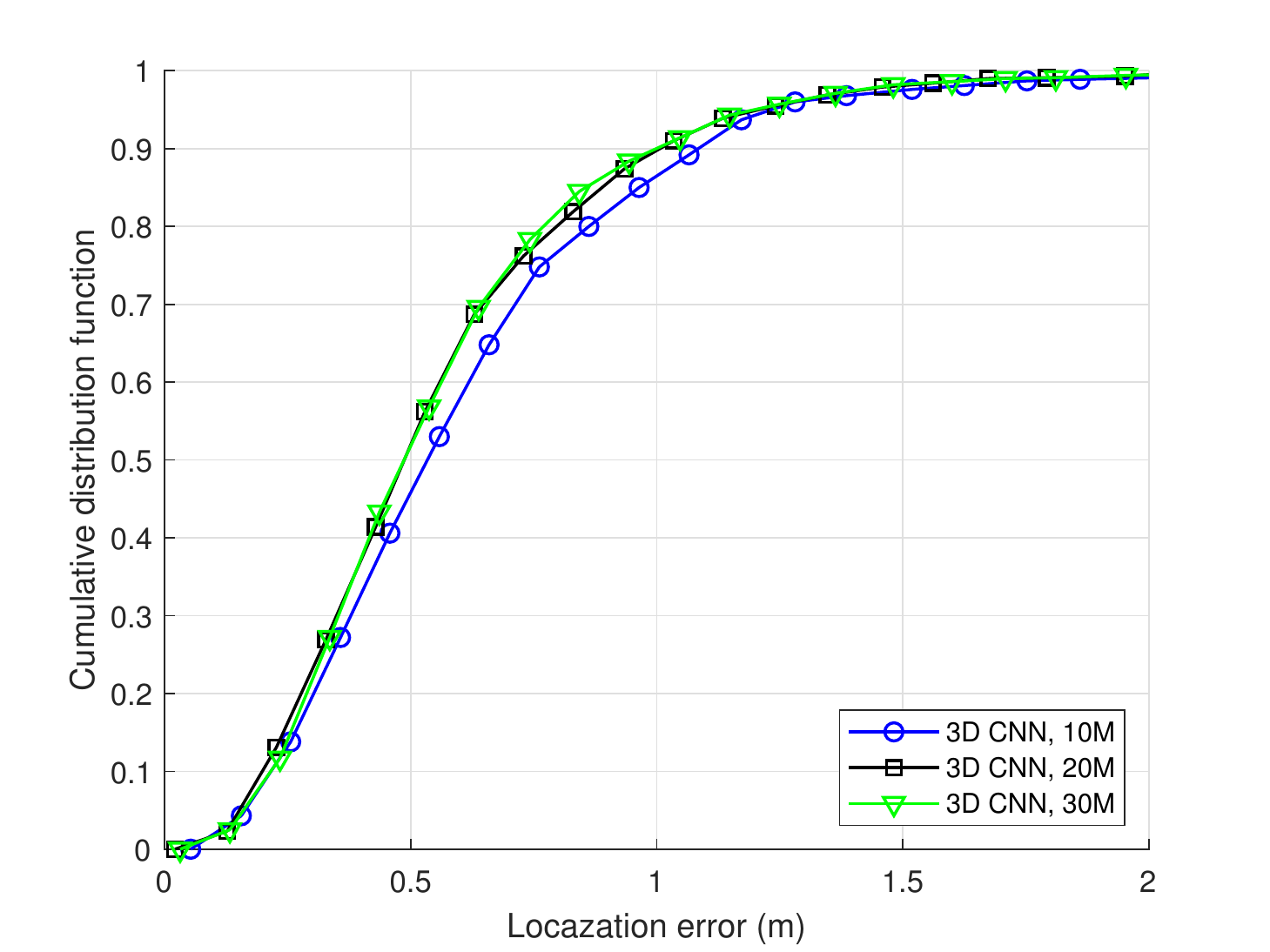}
	\caption{The CDF of localization error with different transmission bandwidth.}
	\label{3DCNNerrorViaBW}
\end{figure}

To inspect how vertical and horizontal angles interact with each other, we compare the different antenna array geometry. Fig. \ref{3DCNNerrorViaAnt} shows the CDF of localization error using different antenna arrangements, where the case of $1\times 128$ corresponds to the uniform linear antenna (ULA) array. To make a fair comparison, we maintain the same number of antenna elements. It can be observed that the two UPAs perform better with 90\% ($8\times16$) and 93\% ($4\times32$) reliability for 1m positioning accuracy, while the ULA performs 80\% reliability for 1m positioning accuracy. Thus, the UPA with a 3D CNN positioning method is more suitable for the 3D positioning than the ULA by providing additional angle resolution in the vertical direction.

Next, we evaluate the impact of the number of antennas. Fig. \ref{3DCNNerrorViaAntNum} demonstrates the CDF of localization error with the number of BS antennas. The number of antennas is increased from 32 to 128 with the BS equipped with UPA. The localization errors at the 90\% point are 1m, 1.05m, and 1.26m for the $8 \times 16$, $4 \times 16$, and $4 \times 8$ UPAs respectively. The results clearly show that the positioning accuracy improves with both the increase of the number of antennas in column and row. Nevertheless, such improvement is getting saturated when the number of antennas reaches 64.

To understand how the delay domain features affect positioning accuracy, we evaluate the performance versus the transmission bandwidth. We consider that the transmission bandwidth rises from 10MHz to 30MHz. From Fig. \ref{3DCNNerrorViaAntNum}, the localization errors at the 90\% point are 1m, 1m, and 1.08m for the 30MHz, 20MHz, and 10MHz transmission bandwidth respectively. It can be observed that when the bandwidth is increased from 10MHz to 20MHz, the positioning accuracy is also growing, while the further increase to 30MHz does not help, indicating that the sufficient delay information (e.g., 20MHz) is able to capture all features in the 3D CNN model.

\subsection{Positioning Robustness}
\begin{figure}[t]
	\centering
	\includegraphics[scale=0.6]{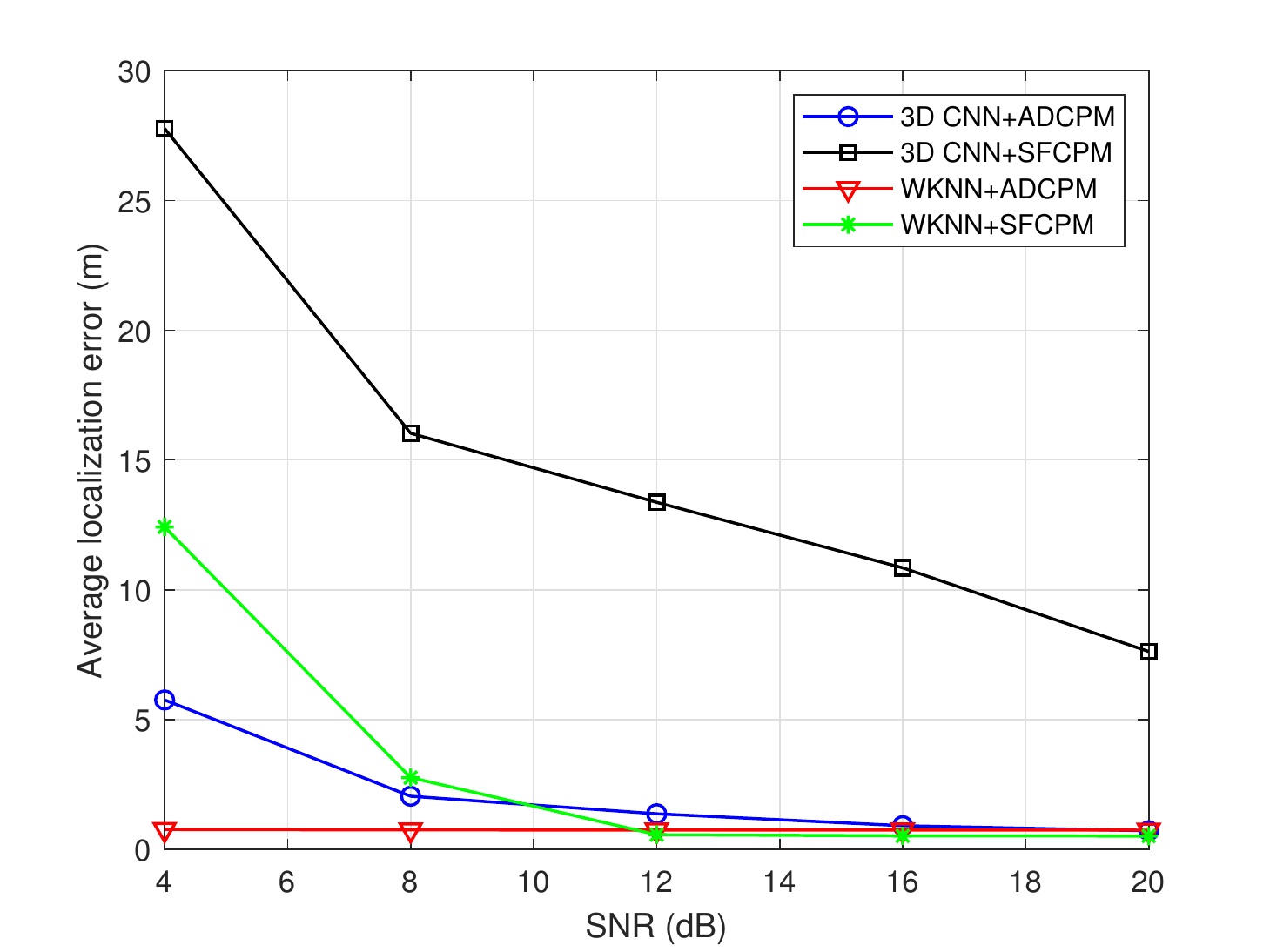}
	\caption{The CDF of localization error with different SNR using ADCPM and SFCPM respectively.}
	\label{errorViaSNR}
\end{figure}
In this subsection, we evaluate the robustness of our fingerprint extraction and 3D CNN-enabled positioning method against noise contamination. For comparison, we also consider the direct use of the information in the space-frequency domain. Similar to the ADCPM, we introduce the space-frequency channel power matrix (SFCPM) of the $k$th MT as
\begin{equation}
{{\mathbf{{{\mathbf{\Xi }}}}}_k} \triangleq E\left\{ {{{\mathbf{H}}_k} \odot {\mathbf{H}}_k^*} \right\}.
\end{equation}
For comparison, the SFCPM is employed as an alternative fingerprint.

Then the following four combinations for the positioning method have been considered: a) WKNN with the SFCPM fingerprint; b) WKNN with the proposed ADCPM fingerprint; c) the proposed 3D CNN with the SFCPM fingerprint; d) the proposed 3D CNN with the proposed ADCPM fingerprint. In the training phase, the artificial noiseless channel matrices are generated by the Monte-Carlo method to train the model for 3D CNN or to construct the look-up table for WKNN. In the prediction phase, real noisy channels are used for positioning, where the average received signal-to-noise ratio (SNR) varies from 4 dB to 20 dB. By exploiting the sparsity of the ADCPM, we adopt a filter with a threshold on the value of the elements of ADCPM such that the values below this threshold is set to 0. In doing so, the noise contamination in the inputs is somewhat reduced.

Fig. \ref{errorViaSNR} presents the average localization error versus the average received SNR. It can be observed that for both the searching-based method (i.e., WKNN) and the learning-based (i.e., 3D CNN) methods, the ADCPM  fingerprint is more robust to noise contamination than its SFCPM counterpart thanks to channel sparsity in the angle-delay domain. At high SNR, while the searching-based method has the comparable performance with both SFCPM and ADCPM fingerprints, the learning-based method favors the ADCPM fingerprint, which introduces sparsity. As the SNR grows, the performance of the learning-based method approaches that of the searching-based one with considerably reduced computational complexity and storage requirement. We conclude that the proposed ADCPM fingerprint is more robust to the noise and the proposed  3D CNN-enabled positioning method performs well for the noisy inputs.

\section{Conclusion}\label{sec5}
In this paper, we have proposed a learning-based user positioning method using 3D CNN for the 3D massive MIMO-OFDM system, exploiting the sparsity properties of channel statistics in the 3D angle-delay domain. We employed the 3D ADCPM as a new type of fingerprint which includes stable and stationary multipath characteristics, e.g. delay, power, and angle in the vertical and horizontal directions. By casting user positioning as an 3D image recognition problem, we proposed a novel 3D CNN architecture with the aim of regression to realize the fingerprint-based positioning. The simulation results demonstrated that the proposed method can achieve high positioning accuracy, reducing the computational complexity and storage overhead, and robust to the noise contamination of the fingerprints.
\begin{appendices}	
	\section{Proof of throrem 1}\label{proof of thro2}
	 In order to prove Theorem 1, we provide the following preliminary results stated as Lemma 1-2.
	 
	 We define the following vectors 
	\begin{equation}\label{MIMO: eq6}
	{\tilde{\bf e}}_{p,k}^{(v)} =\frac{1}{\sqrt M } {\bf{V}}_M^H{{\bf{e}}^{(v)}}({\theta _{p,k}}),
	\end{equation}
	and
	\begin{equation}\label{MIMO: eq7}
	{\tilde {\bf e}}_{p,k}^{(h)} = \frac{1}{\sqrt N }{\bf{V}}_N^H{{\bf{e}}^{(h)}}({\theta _{p,k}},{\varphi _{p,k}}).
	\end{equation}

	\textit{Lemma 1:}
	When $M\to \infty $ and $N\to \infty$, the AOA is extracted from array response vector explicitly via the DFT operation, given by
	\begin{equation}\label{MIMO: eq8}
	\mathop {\lim }\limits_{M \to \infty } {\tilde{ \bf e}}_{p,k}^{(v)} = {\boldsymbol{\alpha }}_M^{({{\bar m}_{p,k}})},
	\end{equation}
	\begin{equation}\label{MIMO: eq9}
	\mathop {\lim }\limits_{N \to \infty } {\tilde{ \bf e}}_{p,k}^{(h)} =  {\boldsymbol{\alpha }}_N^{({{\bar n}_{p,k}})},
	\end{equation}
	where
	\begin{equation}
	{\boldsymbol{\alpha }}_i^{(j)} = {[\underbrace {0, \cdots 0}_{j },1,\underbrace {0, \cdots ,0}_{i - j - 1}]^T},
	\end{equation}

	\textit{Proof:} 	
	The $i$th element of ${\tilde{\mathbf e}}_{p,k}^{(v)}$ is calculated as
	\begin{equation}\label{MIMO: eq15}
	\begin{aligned}
	{[{\tilde{\mathbf e}}_{p,k}^{(v)}]_i} &= \frac{1}{{ \sqrt M }}\sum\limits_{m = 0}^{M - 1} \frac{1}{{ \sqrt M }}{{e^{\bar{\jmath}2\pi \frac{{m(i - M/2)}}{M}}}{e^{ - \bar{\jmath}2\pi m\frac{{d_t^{(v)}}}{{{\lambda _c}}}\cos {\theta _{p,k}}}}}  \\ 
	&= \frac{1}{{ M }}{e^{ - \bar{\jmath}\pi (M - 1)\left( {\frac{{d_t^{(v)}}}{{{\lambda _c}}}\cos {\theta _{p,k}} - \frac{i}{M} + \frac{1}{2}} \right)}}\\
	&\quad\cdot\frac{{\sin \left( {\frac{M}{2}\left( {\frac{{d_t^{(v)}}}{{{\lambda _c}}}\cos {\theta _{p,k}} - \frac{i}{M} + \frac{1}{2}} \right)} \right)}}{{\sin \left( {\frac{1}{2}\left( {\frac{{d_t^{(v)}}}{{{\lambda _c}}}\cos {\theta _{p,k}} - \frac{i}{M} + \frac{1}{2}} \right)} \right)}} \\ 
	&= {e^{ - \bar{\jmath}\pi (M - 1)\left( {\frac{{d_t^{(v)}}}{{{\lambda _c}}}\cos {\theta _{p,k}} - \frac{i}{M} + \frac{1}{2}} \right)}}\\
	&\quad\cdot f_M\left( {\frac{1}{2}\left( {\frac{{d_t^{(v)}}}{{{\lambda _c}}}\cos {\theta _{p,k}} - \frac{i}{M} + \frac{1}{2}} \right)} \right)\\
	\end{aligned} ,
	\end{equation}
	where $f_M(\cdot)$ is defined in (\ref{MIMO: fM}). From (\ref{MIMO: eq15}), ${[{\tilde{\mathbf e}}_{p,k}^{(v)}]_i} \ne 0$ if and only if $f_M\left( {\frac{1}{2}\left( {\frac{{d_t^{(v)}}}{{{\lambda _c}}}\cos {\theta _{p,k}} - \frac{i}{M} + \frac{1}{2}} \right)} \right) \ne 0$. Further, $\mathop {\lim }\limits_{M \to \infty } f_M(x) \ne 0$ if and only if $\sin (x) = 0$, and in this case we have
	\begin{equation}\label{MIMO: eq18}
	\mathop {\lim }\limits_{M \to \infty } f_M(x){|_{x = n\pi ,n = 0, \pm 1, \cdots }} = \frac{{\cos (Mx)}}{{\cos (x)}},
	\end{equation}
	hence, $\mathop {\lim }\limits_{M \to \infty } f_M\left( {\frac{1}{2}\left( {\frac{{d_t^{(v)}}}{{{\lambda _c}}}\cos {\theta _{p,k}} - \frac{i}{M} + \frac{1}{2}} \right)} \right) \ne 0$ if and only if the following constraint is satisfied
	\begin{equation}
	\frac{1}{2}\left( {\frac{{d_t^{(v)}}}{{{\lambda _c}}}\cos {\theta _{p,k}} - \frac{i}{M} + \frac{1}{2}} \right) = n\pi ,n = 0, \pm 1, \cdots .
	\end{equation}
	Owing to $0 \leqslant i \leqslant M-1$, we derive that $\mathop {\lim }\limits_{M \to \infty } {[{\tilde{\mathbf e}}_{p,k}^{(v)}]_i} \ne 0$ if and only if
	\begin{equation}
	i = {{\bar m}_{p,k}} = \frac{M}{2} + \frac{{Md_t^{(v)}}}{{{\lambda _c}}}\cos {\theta _{p,k}},
	\end{equation}
	where in the case ${M \to \infty }$, $\left(\frac{M}{2} + \frac{{Md_t^{(v)}}}{{{\lambda _c}}}\cos {\theta _{p,k}}\right)$ is sufficient to be an integer.
	In this case, we derive from (\ref{MIMO: eq15}) and  (\ref{MIMO: eq18}) that
	\begin{equation}
	\mathop {\lim }\limits_{M \to \infty } {\tilde{\mathbf e}}_{p,k}^{(v)} =  {\boldsymbol {\alpha }}_M^{({{\bar m}_{p,k}})}.
	\end{equation}
	Therefore, (\ref{MIMO: eq8}) is obtained. (\ref{MIMO: eq9}) can be derived by using the same method. \qed
	
	Lemma 1 establishes the relationship between the array response vector and the deterministic component in the angle domain for 3D massive MIMO antennas, which guides us to extract both vertical and horizontal AOA specified by ${{\bar m}_{p,k}}$ and ${{\bar n}_{p,k}}$ respectively in the 3D space. 
	
	Based on Lemma 1, the CIR vector in the angle domain of the $p$th path for the $k$th MT can be written as
	\begin{equation}\label{MIMO: eq5}
	{{{\tilde{\bf q}}}_{p,k}} = \frac{1}{\sqrt {MN}} \left( {{\bf{V}}_M^H \otimes {\bf{V}}_N^H} \right){{\bf{q}}_{p,k}},
	\end{equation}
	in which the index indicates the AOA and the corresponding element indicates the channel gain in the AOA direction.
	
	\textit{Lemma 2:}
	For 3D MIMO systems with UPA at the BS, when $M\to \infty $, the average channel power of each path is concentrated on specific position in the vertical angle domain, given by
	
	\begin{equation}\label{MIMO: eqnew15}
	\begin{aligned}
	\mathop {\lim }\limits_{M \to \infty  } E\left\{ {{{\left| {[{{{\tilde{\mathbf q}}}_{p,k}}]_i} \right|}^2}} \right\} = &\sigma _{p,k}^2f_N^2\left( {\frac{{{{\bar n}_{p,k}} - {\left\langle i \right\rangle }_N  }}{{2N}}} \right)\\ &\cdot \delta \left( {{ \left\lfloor {i/N} \right\rfloor} - {{\bar m}_{p,k}}} \right)
	\end{aligned}.
	\end{equation}
	When $N\to \infty$, the average channel power of each path has the similar conclusion in the horizontal angle domain, given by
	\begin{equation}\label{MIMO: eqnew16}
	\begin{aligned}
	\mathop {\lim }\limits_{N \to \infty } E\left\{ {{{\left| {[{{{\tilde{\mathbf q}}}_{p,k}}]_i} \right|}^2}} \right\} = &\sigma _{p,k}^2f_M^2\left( {\frac{{{{\bar m}_{p,k}} -   \left\lfloor {i/N} \right\rfloor}}{{2M}}} \right)\\
	& \cdot \delta \left( {  {{\left\langle i \right\rangle }_N} - {{\bar n}_{p,k}}} \right)
	\end{aligned}.
	\end{equation}
	When $M\to \infty $ and $N\to \infty$, the average channel power of each path is concentrated in both the vertical and horizontal angle domain, given by
	\begin{equation}\label{MIMO: eqnew17}
	\begin{aligned}
	\mathop {\lim }\limits_{M \to \infty ,N \to \infty } E\left\{ {{{\left| {{{[{{{\tilde{\mathbf q}}}_{p,k}}]}_i}} \right|}^2}} \right\} =  
	\sigma _{p,k}^2 \delta \left( i- {{\bar m}_{p,k}}N - {{\bar n}_{p,k}} \right)
	\end{aligned}  .
	\end{equation}
	
	\textit{Proof:} Substituting (\ref{MIMO: eq1}) and (\ref{MIMO: eq2}) into (\ref{MIMO: eq5}), we have
	\begin{equation}\label{MIMO: eq21}
	\begin{aligned}
	{{{\tilde{\mathbf q}}}_{p,k}} &= \frac{1}{\sqrt {MN}}{a_{p,k}}\left( {{\mathbf{V}}_M^H \otimes {\mathbf{V}}_N^H} \right)\\
	& \quad \cdot \left( {{{\mathbf{e}}^{(v)}}({\theta _{p,k}}) \otimes {{\mathbf{e}}^{(h)}}({\theta _{p,k}},{\varphi _{p,k}})} \right) \hfill \\
	&= {a_{p,k}}\left( {\frac{1}{\sqrt {M}}{\mathbf{V}}_M^H{{\mathbf{e}}^{(v)}}({\theta _{p,k}})} \right) \\
	& \quad \otimes \left(\frac{1}{\sqrt {N}} {{\mathbf{V}}_N^H{{\mathbf{e}}^{(h)}}({\theta _{p,k}},{\varphi _{p,k}})} \right) \hfill \\ 
	&={a_{p,k}}{\tilde{\mathbf e}}_{p,k}^{(v)} \otimes {\tilde{\mathbf e}}_{p,k}^{(h)} \hfill \\ 
	\end{aligned} ,
	\end{equation}
	when $M\to \infty $, substituting (\ref{MIMO: eq8}) into (\ref{MIMO: eq21}), we have
	\begin{equation}
	\mathop {\lim }\limits_{M \to \infty} {{{\tilde{\mathbf q}}}_{p,k}} = {a_{p,k}}{\boldsymbol {\alpha }}_M^{({{\bar m}_{p,k}})} \otimes {\tilde{\mathbf e}}_{p,k}^{(h)}.
	\end{equation}
	Similar to (\ref{MIMO: eq15}), the $i$th element of ${\tilde{\mathbf e}}_{p,k}^{(h)}$ is calculated as
	\begin{equation}\label{MIMO: eq25}
	\begin{aligned}
	\mathop {\lim }\limits_{M \to \infty }{[{\tilde{\mathbf e}}_{p,k}^{(h)}]_i} =  {e^{ - \bar{\jmath}\pi (N - 1)\left( {\frac{{{{\bar n}_{p,k}} - i}}{N}} \right)}}{f_N}\left( {\frac{{{{\bar n}_{p,k}} - i}}{{2N}}} \right) \hfill \\ 
	\end{aligned}. 
	\end{equation}
	Therefore, the $i$th element of ${{\tilde{\mathbf q}}_{p,k}}$ is written as
	\begin{equation} 
	\begin{aligned}
	\mathop {\lim }\limits_{M \to \infty }{[{{{\tilde{\mathbf q}}}_{p,k}}]_i} = 
	&{a_{p,k}}{e^{ - \bar{\jmath}\pi (N - 1)\left( {\frac{{{{\bar n}_{p,k}} - {{\left\langle i \right\rangle }_N} }}{N}} \right)}}\\
	&\cdot{f_N}\left( {\frac{{{{\bar n}_{p,k}} - {{\left\langle i \right\rangle }_N}}}{{2N}}} \right)\delta \left( {\left\lfloor {i/N} \right\rfloor  - {{\bar m}_{p,k}}} \right)
	\end{aligned}.
	\end{equation}
	From the assumption that ${a_{p,k}}\sim \mathcal{CN}(0,\sigma _{p,k}^2)$, the average channel power in angle domain is calculated as
	\begin{equation}
	\begin{aligned}
	\mathop {\lim }\limits_{M \to \infty } E\left\{ {{{\left| {[{{{\tilde{\mathbf q}}}_{p,k}}]_i} \right|}^2}} \right\} = &\sigma _{p,k}^2f_N^2\left( {\frac{{{{\bar n}_{p,k}} - {{\left\langle i \right\rangle }_N} }}{{2N}}} \right)\\ &\cdot \delta \left( { \left\lfloor {i/N} \right\rfloor - {{\bar m}_{p,k}}} \right)
	\end{aligned}.
	\end{equation}
	
	Therefore, (\ref{MIMO: eqnew15}) is obtained. When $N\to \infty$, (\ref{MIMO: eqnew16}) can be derived by using the same method.
	
	When $M\to \infty $ and $N\to \infty$, (\ref{MIMO: eqnew17}) is a special case of (\ref{MIMO: eqnew15}) and (\ref{MIMO: eqnew16}). Substituting (\ref{MIMO: eq8}) and (\ref{MIMO: eq9}) into (\ref{MIMO: eq21}), we derive
	\begin{equation}
	\mathop {\lim }\limits_{M \to \infty ,N \to \infty } {{{\tilde{\mathbf q}}}_{p,k}} = {a_{p,k}} {\mathbf{\alpha }}_M^{({{\bar m}_{p,k}})} \otimes {\mathbf{\alpha }}_N^{({{\bar n}_{p,k}})}.
	\end{equation}
	Hence, the $i$th element of ${{\tilde{\mathbf q}}_{p,k}}$ can be written as
	\begin{equation}  
	\begin{aligned}
	\mathop {\lim }\limits_{M \to \infty ,N \to \infty }{[{{{\tilde{\mathbf q}}}_{p,k}}]_i} =
	{a_{p,k}} \delta (i -{{\bar m}_{p,k}} N - {{\bar n}_{p,k}} )
	\end{aligned}  .
	\end{equation}
	From the assumption that ${a_{p,k}}\sim \mathcal{CN}(0,\sigma _{p,k}^2)$, the average channel power in angle domain is calculated as
	\begin{equation}
	\begin{aligned}
	\mathop {\lim }\limits_{M \to \infty ,N \to \infty } E\left\{ {{{\left| {{{[{{{\tilde{\mathbf q}}}_{p,k}}]}_i}} \right|}^2}} \right\} =  
	\sigma _{p,k}^2 \delta \left( i- {{\bar m}_{p,k}}N - {{\bar n}_{p,k}} \right)
	\end{aligned}  .
	\end{equation}
	Therefore, (\ref{MIMO: eqnew17}) is obtained. \qed
	
	Lemma 2 establishes the relationship between the CIR in the antenna domain and that in the angle domain for 3D massive MIMO systems. When the number of UPA's antennas is sufficiently large, the average channel powers corresponding to different AOA can be resolved via DFT operation. The resolution of power and angle in the vertical and horizontal directions depends on the number of antennas in the vertical and horizontal directions respectively.

	The $j$th column of $\frac{1}{\sqrt{N_c}} {{\mathbf{H}}_k}{\mathbf{F}}_{{N_c} \times {N_g}}^*$ is calculated as
	\begin{equation}
	\begin{aligned}
	&\frac{1}{\sqrt{N_c}}{[{{\mathbf{H}}_k}{\mathbf{F}}_{{N_c} \times {N_g}}^*]_j}\\ &= \frac{1}{{\sqrt {{N_c}} }}\sum\limits_{m = 0}^{{N_c} - 1} {\sum\limits_{p = 1}^{N_p} {{a_{p,k}}{\mathbf{e}}({\theta _{p,k}},{\varphi _{p,k}}){e^{ - \bar{\jmath}2\pi \frac{{m{r_{p,k}}}}{{{N_c}}}}}\frac{1}{{\sqrt {{N_c}} }}{e^{ - \bar{\jmath}2\pi \frac{{mj}}{{{N_c}}}}}} }  \hfill \\
	&= \frac{1}{{{{N_c}} }}\sum\limits_{p = 1}^{N_p} {{a_{p,k}}{\mathbf{e}}({\theta _{p,k}},{\varphi _{p,k}})\sum\limits_{m = 0}^{{N_c} - 1} {{e^{ - \bar{\jmath}2\pi \frac{{m({r_{p,k}} - j)}}{{{N_c}}}}}} }  \hfill \\
	&= \frac{1}{{\ {{N_c}} }}\sum\limits_{p = 1}^{N_p} {a_{p,k}}{\mathbf{e}}({\theta _{p,k}},{\varphi _{p,k}}){e^{ - \bar{\jmath}\pi ({N_c} - 1)}} \frac{{\sin \left( {\frac{1}{2}({r_{p,k}} - j)} \right)}}{{\sin \left( {\frac{1}{{2{N_c}}}({r_{p,k}} - j)} \right)}}  \hfill \\ 
	&= \sum_{p=1}^{N_p} a_{p, k} \mathbf{e}\left(\theta_{p, k}, \varphi_{p, k}\right) e^{-\bar{\jmath} \pi\left(N_{c}-1\right)}f_{N_{c}}\left(\frac{r_{p, k}-j}{2 N_{c}}\right) \hfill\\
	\end{aligned} ,
	\end{equation}
	where $f_{N_c}(\cdot)$ is defined by substituting $N_c$ for $M$ in (\ref{MIMO: fM}). When $N_c\to \infty$, $\mathop {\lim }\limits_{{N_c} \to \infty } f_{N_{c}}\left(\frac{1}{2 N_{c}}\left(r_{p, k}-j\right)\right) \ne 0$ if and only if $j = {r_{p,k}}$. In this case, from (\ref{MIMO: eq18}), we have
	\begin{equation}
	\begin{aligned}
	{\frac{1}{\sqrt{N_c}}[{{\mathbf{H}}_k}{\mathbf{F}}_{{N_c} \times {N_g}}^*]_j} = & \sum\limits_{p = 1}^{N_p} {{a_{p,k}}{\mathbf{e}}({\theta _{p,k}},{\varphi _{p,k}}){e^{ - j\pi ({N_c} - 1)}} }\\&\cdot \delta(j - {r_{p,k}})
	\end{aligned}.
	\end{equation}
	Therefore, we derive
	\begin{equation}\label{MIMO: eq30}
	\mathop {\lim }\limits_{{N_c} \to \infty }{\frac{1}{\sqrt{N_c}}{\mathbf{H}}_k}{\mathbf{F}}_{{N_c} \times {N_g}}^* =  \sum\limits_{p = 1}^{N_p} {{a_{p,k}}{\mathbf{e}}({\theta _{p,k}},{\varphi _{p,k}})} {\left[ {{\mathbf{\alpha }}_{{N_g}}^{({r_{p,k}})}} \right]^T}.
	\end{equation}
	Substituting (\ref{MIMO: eq2}) (\ref{MIMO: eq6}), (\ref{MIMO: eq7}) and (\ref{MIMO: eq30}) into (\ref{MIMO: eq26}), we have
	\begin{equation}
	\begin{aligned}
	\mathop {\lim }\limits_{{N_c} \to \infty }{{\mathbf{G}}_k}&= \frac{1}{\sqrt {MN}} \sum\limits_{p = 1}^{N_p} {{a_{p,k}}\left( {{\mathbf{V}}_M^H \otimes {\mathbf{V}}_N^H} \right){\mathbf{e}}({\theta _{p,k}},{\varphi _{p,k}})} {\left[ {{\mathbf{\alpha }}_{{N_g}}^{({r_{p,k}})}} \right]^T} \hfill \\
	&= \frac{1}{\sqrt {MN}} \sum\limits_{p = 1}^{N_p} {{a_{p,k}}} \left( {{\mathbf{V}}_M^H \otimes {\mathbf{V}}_N^H} \right)\\
	&\quad\cdot\left( {{{\mathbf{e}}^{(v)}}({\theta _{p,k}}) \otimes {{\mathbf{e}}^{(h)}}({\theta _{p,k}},{\varphi _{p,k}})} \right){\left[ {{\mathbf{\alpha }}_{{N_g}}^{({r_{p,k}})}} \right]^T} \hfill \\
	&=  \sum\limits_{p = 1}^{N_p} {{a_{p,k}}} \left( \frac{1}{\sqrt {M}}{{\mathbf{V}}_M^H{{\mathbf{e}}^{(v)}}({\theta _{p,k}})} \right)  \\
	&\quad\otimes\left( \frac{1}{\sqrt {N}}{{\mathbf{V}}_N^H{{\mathbf{e}}^{(h)}}({\theta _{p,k}},{\varphi _{p,k}})} \right){\left[ {{\mathbf{\alpha }}_{{N_g}}^{({r_{p,k}})}} \right]^T} \hfill \\
	&= \sum\limits_{p = 1}^{N_p} {{a_{p,k}}} \left( {{\tilde{\mathbf e}}_{p,k}^{(v)} \otimes {\tilde{\mathbf e}}_{p,k}^{(h)}} \right){\left[ {{\mathbf{\alpha }}_{{N_g}}^{({r_{p,k}})}} \right]^T} \hfill \\
	\end{aligned}.
	\end{equation}
	
	When $M\to \infty $ and $N_c\to \infty$, according to (\ref{MIMO: eq8}) and (\ref{MIMO: eq25}), we have
	\begin{equation}
	\begin{aligned}
	\mathop {\lim }\limits_{M \to \infty,{N_c} \to \infty } {[{\mathbf{G}}_k]_{i,j}}= & \sum\limits_{p = 1}^{N_p} {a_{p,k}}{f_N} \left( {\frac{{{{\bar n}_{p,k}} - {\left\langle i \right\rangle }_N}}{{2N}}} \right)\\&\quad \cdot \delta \left( {{  \left\lfloor {i/N} \right\rfloor  } - {{\bar m}_{p,k}}} \right) \delta \left( {j - {r_{p,k}}} \right)
	\end{aligned}.
	\end{equation}
	From the assumption that ${a_{p,k}}\sim \mathcal{CN}(0,\sigma _{p,k}^2)$, the average channel power in angle-delay domain is calculated as
	\begin{equation}\label{MIMO: eq79}
	\begin{aligned}
	\mathop {\lim }\limits_{M \to \infty,{N_c} \to \infty } &E\left\{ {{{\left| {{{[{\bf G}_k]}_{i,j}}} \right|}^2}} \right\} \\&= MN{N_c}\sum\limits_{p = 1}^{N_p} \sigma _{p,k}^2f_N^2\left( {\frac{{{{\bar n}_{p,k}} - {\left\langle i \right\rangle }_N }}{{2N}}} \right)\\&\quad \cdot \delta \left( {{ \left\lfloor {i/N} \right\rfloor } - {{\bar m}_{p,k}}} \right) \delta \left( {j - {r_{p,k}}} \right)
	\end{aligned}.
	\end{equation}
	Substituting (\ref{MIMO: eq31}) into (\ref{MIMO: eq79}), (\ref{MIMO: eq34}) is obtained.
	
	When $M\to \infty $, and $N_c\to \infty$, (\ref{MIMO: eq35}) can be derived by using the same method.

	When $M\to \infty $, $N\to \infty$ and $N_c\to \infty$, according to (\ref{MIMO: eq8}) and (\ref{MIMO: eq9}), we have
	\begin{equation}
	\begin{aligned}
	\mathop {\lim }\limits_{M \to \infty ,N \to \infty ,{N_c} \to \infty } {{\mathbf{G}}_k}= & \sum\limits_{p = 1}^{N_p} {{a_{p,k}}} 
	 \left( {{\boldsymbol{\alpha }}_M^{({{\bar m}_{p,k}})} \otimes {\boldsymbol{\alpha }}_N^{({{\bar n}_{p,k}})}} \right)\\&\cdot{\left[ {{\boldsymbol{\alpha }}_{{N_g}}^{({r_{p,k}})}} \right]^T}
	\end{aligned}.
	\end{equation}
	From the assumption that ${a_{p,k}}\sim \mathcal{CN}(0,\sigma _{p,k}^2)$, the average channel power in angle-delay domain is calculated as
	\begin{equation}\label{MIMO: eq82}
	\begin{aligned}
	\mathop {\lim }\limits_{M \to \infty ,N \to \infty ,{N_c} \to \infty } &E\left\{ {{{\left| {{{[{{\mathbf{G}}_k}]}_{i,j}}} \right|}^2}} \right\} 
	\\&= \sum\limits_{p = 1}^{N_p} \sigma _{p,k}^2\delta \left( {i - {{\bar m}_{p,k}} N - {{\bar n}_{p,k}}} \right)\\
	&\quad\cdot\delta \left( {j - {r_{p,k}}} \right)
	\end{aligned}.
	\end{equation}
	Substituting (\ref{MIMO: eq31}) into (\ref{MIMO: eq82}), we get (\ref{MIMO: eq36}).
\end{appendices}

\end{document}